\documentclass[preprint,eqsecnum,showpacs,nofootinbib,aps,amsmath,amssymb,tightenlines]{revtex4}

\usepackage{graphicx}

\newcommand{\Lie}[0]{{\cal L}\, }
\newcommand{\grad}[0]{\nabla\!}
\newcommand{\R}{{\mathcal{R}}}
\def\th{{\widehat{\tau}}}
\def\rh{{\widehat{r}}}
\def\twoR{{\widetilde{\R}}}
\def\ls{{(\ell)}}
\def\ns{{(n)}}
\def\l{{\ell}}
\def\k{\bar{\kappa}}
\def\fie{\varphi}
\def\q{{\widetilde{q}}}
\def\K{{\widetilde{K}}}
\def\lb{\bar{\ell}}
\def\nb{\bar{n}}
\def\Vb{\bar{V}}
\newcommand{\man}{{\mathcal{M}}}
\def\b{$\bullet$}
\def\T{\bar{T}}
\def\m{{\rm matter}}
\def\g{{\rm grav}}
\def\twoD{\widetilde{D}}
\def\W{\widetilde{W}}
\def\S{\widetilde{S}}
\newcommand{\G}{{\mathcal{G}}}
\def\Mdot{{\dot{M}}}

\def\be{\begin{equation}}
\def\ee{\end{equation}}
\def\ba{\begin{eqnarray}}
\def\ea{\end{eqnarray}}

\preprint{\vbox{\baselineskip=12pt \rightline{gr-qc/0308033}
\rightline{CGPG-03-07/3} \rightline{NSF-KITP-03-57}}}

\begin{document}

\title{Dynamical Horizons and their properties}
\date{\today}
    \author{Abhay Ashtekar${}^{1,2,4}$}
   \email{ashtekar@gravity.psu.edu}
    \author{Badri Krishnan${}^{1,3,4}$}
    \email{badkri@aei.mpg.de}
    \affiliation{${}^1$Center for Gravitational Physics and Geometry,\\
    Physics Department, Penn State University, \\University Park, PA 16802,
    USA\\
    ${}^2$Kavli Institute of Theoretical Physics\\
    University of California, Santa Barbara, CA 93106-4030, USA\\
    ${}^3$Max Planck Institut f\"ur Gravitationsphysik\\
    Albert Einstein Institut, 14476 Golm, Germany\\
    ${}^4$Erwin Schr\"odinger Institute, Boltzmanngasse 9, 1090 Vienna,
    Austria\\}

\begin{abstract}

A detailed description of how black holes grow in full, non-linear
general relativity is presented. The starting point is the notion
of \emph{dynamical horizons}. Expressions of fluxes of energy and
angular momentum carried by gravitational waves across these
horizons are obtained. Fluxes are local and the energy flux is
positive. Change in the horizon area is related to these fluxes. A
notion of angular momentum and energy is associated with
cross-sections of the horizon and balance equations, analogous to
those obtained by Bondi and Sachs at null infinity, are derived.
These in turn lead to generalizations of the first and second laws
of black hole mechanics. The relation between dynamical horizons
and their asymptotic states ---the isolated horizons--- is
discussed briefly. The framework has potential applications to
numerical, mathematical, astrophysical and quantum general
relativity.

\end{abstract}


\maketitle

\section{Introduction}
\label{s1}

Properties of stationary, 4-dimensional  black holes have been
well-understood for quite some time. In the Einstein-Maxwell
theory, for example, the situation is astonishingly simple: We
know that there is a \emph{unique} 4-parameter family of
stationary solutions and, furthermore, these solutions are known
explicitly, in a closed form, given by the Kerr-Newman metrics and
associated Maxwell fields \cite{mh}. Large families of stationary
but distorted black holes are also known, where the distortion is
caused by rings of matter and magnetic fields \cite{dis}. Finally,
a framework has recently been introduced to probe properties of
black holes which are themselves in equilibrium but in space-times
with non-trivial dynamics in the exterior region
\cite{abf,ihprl,abl1}. In particular, this \emph{isolated horizon}
framework enables one to assign mass and angular momentum to black
holes in terms of values of the fields on the horizon itself,
without any reference to infinity and has also led to a
generalization of the zeroth and first laws of black hole
mechanics \cite{afk,abl2}.

However, in Nature, black holes are rarely in equilibrium. They
grow by swallowing stars and galactic debris as well as
electromagnetic and gravitational radiation. For such fully
dynamical black holes, essentially there has been only one major
result in \emph{exact} general relativity. This is the celebrated
area theorem, proved by Stephen Hawking in the early seventies
\cite{swh,he}: if matter satisfies the dominant energy condition,
the area of the black hole event horizon can never decrease. This
theorem has been extremely influential because of its similarity
with the second law of thermodynamics. However, it is a
qualitative result; it does not provide an explicit formula for
the amount by which the area increases in physical situations.
Now, the first law of black hole mechanics
\be \label{1law1} \delta E = (\kappa/8\pi G) \delta a + \Omega
\delta J\, \ee
does relate the change in the area of an isolated horizon to that
in the energy and angular momentum, as the black hole makes a
transition from one equilibrium state to a nearby one. This
suggests that there may well be a fully dynamical version of
(\ref{1law1}) which relates the change in the black hole area to
the energy and angular momentum fluxes, as the black hole makes a
transition from a given state to one which is far removed. Thus,
we are naturally led to ask: Can the results obtained in the
isolated horizon framework be extended to fully dynamical
situations?

Attractive as this possibility seems, one immediately encounters a
serious conceptual and technical problem. For, the expression
requires, in particular, a precise notion of the flux of
gravitational energy across the horizon. Already at null infinity,
the expression of the gravitational energy flux is subtle: one
needs the framework developed by Bondi, Sachs, Newman, Penrose and
others to introduce a viable, gauge invariant expression of this
flux \cite{bondiflux,as,wz}. In the strong field regime, there is
no satisfactory generalization of this framework and no
satisfactory, gauge invariant notion of gravitational energy flux
beyond perturbation theory. Thus, one appears to be stuck right at
the start.

Yet, there are at least two general considerations that suggest
that an extension of the first law to fully dynamical situations
should be possible. Consider a stellar collapse leading to the
formation of a black hole. At the end of the process, one has a
black hole and, from general physical considerations, one expects
that the energy in the final black hole should equal the total
matter plus gravitational energy that fell across the horizon.
Thus, at least the total integrated flux across the horizon should
be well defined. Indeed, it should equal the depletion of the
energy in the asymptotic region, i.e., the difference between the
ADM energy and the energy radiated across future null infinity.
The second consideration involves the Penrose inequalities which
were motivated by cosmic censorship: The ADM mass should be
greater than or equal to the half the radius of the apparent
horizon on any (partial) Cauchy slice \cite{rp}. (Special cases of
this conjecture have been proved recently \cite{hib}.)
Heuristically, the inequality leads us to think of the apparent
horizon radius as a measure of the mass in its interior, whence
one is led to conclude that the change in the area is due to
influx of energy. Thus, it is tempting to hope that something
special may happen at the surface of a black hole enabling one to
define the flux of energy and angular momentum across it, thereby
giving a precise meaning to these physical expectations.

The question then is: how should we define the surface of the
black hole? The obvious candidate is the event horizon.
Unfortunately, this is not a viable possibility because event
horizons are extremely global and teleological. Consider for
example the gravitational collapse of a thin spherical shell. The
event horizon first forms in the interior of the shell and then
expands out. Thus, in the initial phase, it lies in a flat
space-time region and expands out \emph{in anticipation that the
shell will cross it}, even though neither the matter nor the
gravitational radiation falls across it before it hits the shell.
Thus, one cannot hope to find a quasi-local, fully dynamical
generalization of the first law using event horizons. However,
there is an alternative, suggested by the strategy used routinely
in numerical simulations of black hole formation or coalescence.
There, one avoids the problems associated with the global and
teleological nature of the event horizon by locating apparent
horizons at each time during evolution.%
\footnote{In this paper, the term `apparent horizon' is used in
the sense employed in snumerical relativity: it is the outermost
marginally trapped surface on a given (partial) Cauchy slice. By
contrast, Hawking and Ellis \cite{he} define an apparent horizon
as the boundary of a trapped region associated with the Cauchy
slice (i.e., of a connected region through each point of which
there passes an outer trapped surface lying in the slice.}
Can one then use apparent horizons to obtain the desired
generalization of eq. (\ref{1law1})? Now, apparent horizons can
and do jump during evolution. However, in all numerical
simulations, there are epochs during which the world tube
$\tau_{\rm AH}$ traced out by apparent horizons is smooth. The
rough, intuitive idea is to use these world tubes as the black
hole surfaces across which energy and angular momentum fluxes are
to be calculated.

We will incorporate these heuristics in a precise notion called
\emph{dynamical horizons}. However, the definition will only
involve conditions on a 3-surface $H$, extracted from the expected
properties of $\tau_{\rm AH}$. In particular, the definition will
not make any reference to space-time foliations and apparent
horizons thereon. Indeed, the definition will be quasi-local.
Thus, given a region of space-time, one can tell whether or not it
admits dynamical horizons, without any knowledge of the geometry
and matter fields in the exterior region. Similarly, given a
specific 3-dimensional sub-manifold, one can decide whether it is
a dynamical horizon by examining space-time fields defined on it,
without the knowledge of geometry and matter fields away from the
surface. By construction, the world tubes $\tau_{\rm AH}$ will
provide examples of dynamical horizons which are most useful to
numerical relativity. However, using Hayward's \cite{sh} notion of
trapping boundaries, one can also associate with a generic
evolving black hole a more invariantly defined or \emph{canonical}
dynamical horizon. From a general conceptual viewpoint, it may
seem more natural to restrict oneself just to these canonical
dynamical horizons. However, for `practical' applications, this
would be too restrictive. For, although these horizons do not
refer to global notions such as null infinity, they are
nonetheless difficult to locate in a given space-time. A key
strength of the approach is that our analysis is not tied just to
them but encompasses \emph{all} dynamical horizons. In particular,
we will be able to introduce flux formulas and an integral
generalization of the first law (\ref{1law1}) which will hold on
\emph{all} dynamical horizons, including the ones of interest to
numerical relativity.

The paper is organized as follows. In section \ref{s2}, we
introduce the main definitions, motivate the conditions and
explain the relation to Hayward's trapping horizons.  In section
\ref{s3} we derive an \emph{area balance law}, relating the change in
the area of the dynamical horizon to the flux of matter energy and
a pure geometrical, positive definite term. We then interpret the
geometrical term as the flux of gravitational energy and show that
it satisfies the criteria one normally uses to establish the
viability of the Bondi flux formula at null infinity. Section
\ref{s4} introduces the notion of angular momentum and section
\ref{s5} extends the area balance law using angular momentum
considerations to an integral form of the first law.  Using
strategies that have been successful in the isolated horizon
framework, we also introduce a definition of horizon energy and
show that it matches well with the flux formulas to provide an
energy balance law analogous to that at null infinity, but now in
the strong field regime of dynamical horizons. While the horizon
would be dynamical in the time dependent phase of black hole
formation or soon after two black holes merge, one expects it to
settle down and reach equilibrium at late times. Thus, one would
expect isolated horizons to be the asymptotic states of dynamical
horizons. In section \ref{s6} we explore the relation between the
two. Section \ref{s7} summarizes the overall situation, suggests
applications of dynamical horizons to numerical, mathematical and
quantum relativity and lists problems in these areas whose
resolution would shed much new light on how black holes grow and
settle down to their final states.

To preserve the flow of discussion in the main paper, some issues
have been postponed to appendices. Appendix
\ref{a1} discusses the simplest explicit examples of dynamical
horizons and their passage to equilibrium.
For completeness, in Appendix \ref{a3} we discuss
the time-like analogs of dynamical horizons which arise in
cosmological contexts.

The main results of this work were briefly reported in
\cite{ak1,pleban}. Here we present the details, proofs and
extensions of those results.

\section{Definitions and the method}
\label{s2}

In this section, we will introduce the basic definitions, explain
in some detail the motivation behind them, discuss the relation
between dynamical horizons and closely related notions of trapping
horizons introduced by Hayward \cite{sh}, and outline the main
idea on which calculations in the subsequent sections are based.

\subsection{Definition and motivation}
\label{s2.1}

\noindent \textbf{Definition 1:} A smooth, three-dimensional,
space-like sub-manifold $H$ in a space-time $\man$ is said to be a
\emph{dynamical horizon} if it can be foliated by a family of
closed 2-surfaces such that, on each leaf $S$, the expansion
$\Theta_{(\ell)}$ of one null normal $\ell^a$ vanishes and the
expansion $\Theta_{(n)}$ of the other null normal $n^a$ is
strictly negative.%
\footnote{This notion of dynamical horizons is slightly more
general than that used in the brief reports \cite{ak1,pleban}
where the foliation was fixed and the topology of the leaves of the
foliation was required to be $S^2$.}

Thus, basically a dynamical horizon $H$ is a space-like 3-manifold
which is foliated by closed, marginally trapped 2-surfaces. Note
first that, in contrast to event horizons, dynamical horizons can
be located quasi-locally; knowledge of full space-time is not
required. Thus, for example, while an event horizon may well be
developing in the room in which you are now sitting \emph{in
anticipation of a future gravitational collapse}, you can rest
assured that \emph{no} dynamical horizon has ever developed in
that room! Next, since event horizons are defined as the future
boundary of the causal past of future null infinity, the notion is
tied to asymptotically flat space-times. Being quasi-local, the
notion of dynamical horizons does not refer to the asymptotic
structure at all and is meaningful \emph{also in spatially compact
space-times}. On the other hand, while in asymptotically flat
space-times black holes are \emph{characterized} by event
horizons, there is no one-to-one correspondence between black
holes and dynamical horizons. First of all, we expect that
\emph{stationary} black holes do not admit dynamical horizons
because these space-times are non-dynamical. In time dependent
situations, if the dominant energy condition holds and the
space-time is asymptotically predictable, dynamical horizons lie
inside the event horizon. However, in the interior of an expanding
event horizon, there may be many dynamical horizons. Nonetheless,
in the sense made precise in section \ref{s2.2}, under fairly
general conditions one can associate with each evolving black hole
an outermost or canonical dynamical horizon. For conceptual
reasons, it is natural to focus just on this canonical one.
However, our results will apply to \emph{all} dynamical horizons;
indeed, it is this fact that makes the framework powerful in
practice, e.g., for applications to numerical relativity.

Apart from the requirement that $H$ be foliated by marginally
trapped surfaces, the definition contains three conditions. The
first asks that the 2-surfaces which constitute the leaves of the
foliation be closed. This condition is necessary to ensure the
convergence of various integrals we will perform. The second asks
that the expansion $\Theta_{(n)}$ be strictly negative. This
condition is quite weak because, in essence, it simply enables one
to identify $n^a$ as the inward pointing null normal. Thus, had
$\Theta_{(n)}$ been positive, we would be in the white hole
situation, rather than the black hole one. Nonetheless, the
condition \emph{is} restrictive in a minor way: it rules out the
degenerate case in which $\Theta_{(n)}$ vanishes. As we will show
below, the area of the trapped surfaces increases if
$\Theta_{(n)}$ is negative and remains constant if it vanishes.
Thus, by removing the degenerate case, we are basically ignoring
the non-dynamical situation. One might consider intermediate
dynamical situations in which $\Theta_{(n)}$ vanishes on a portion
of each marginally trapped surface and is negative elsewhere. In
this case, the total area would still increase. Our main results
will continue to be valid in these intermediate cases.

The third condition is that $H$ be space-like. Intuitively, it is
clear that if $H$ were time-like, it would not be a boundary of a
black hole region because light rays originating on $H$ would
propagate on both sides of the space-time separated by $H$. So,
the non-triviality lies in the fact that this condition rules out
the possibility that $H$ could be null. To probe how much of a
restriction this is physically, let us proceed by dropping the
requirement that $H$ be space-like but keeping  the other
conditions in Definition 1. Denote by $V^a$ a vector field which
is tangential to $H$, everywhere orthogonal to the foliation by
marginally trapped surfaces and preserves this foliation. We can
always choose the normalization of $\ell^a$ and $n^a$ such that
$\ell^a n_a = -2$ and $V^a = \ell^a - f n^a$ for some $f$. Since
$V\cdot V = 4f$, it follows that $H$ is respectively, space-like,
null or time-like, depending on whether $f$ is positive, zero or
negative. We will argue that under conditions that capture the
physics we have in mind, generically $f$ would be non-negative.
Let us begin by noting that the definition of $V^a$ immediately
implies ${\cal L}_{V}\, \Theta_{(\ell)}= 0$, whence, ${\cal
L}_{\ell}\, \Theta_{(\ell)} = f {\cal L}_{n}\, \Theta_{(\ell)}$.
Therefore, the Raychaudhuri equation for $\ell^a$ implies
\be f\,\Lie_{n}\, \Theta_{(\ell)} = \label{spacelike} -\sigma^2 -
R_{ab}\ell^a\ell^b   \ee
where $\sigma$ is the shear of $\ell^a$. Now, given the scenario
we have in mind, it is physically reasonable to assume that the
convergence $\Theta_{(\ell)}$ of $\ell^a$ becomes negative as one
moves along $n^a$ to the interior of the marginally trapped
surfaces, whence ${\cal L}_{n}\, \Theta_{(\ell)} <0 $. If matter
satisfies the dominant energy condition, the right side of
(\ref{spacelike}) is non-positive, whence we conclude that $f$ is
non-negative; as expected the time-like case is ruled out.
Finally, as we will show in section \ref{s3}, if the flux of
energy across $H$ is non-zero on any one leaf of the foliation of
$H$, the right side of (\ref{spacelike}) cannot vanish identically
on that leaf. Thus, under the intended dynamical situations, $f$
would be strictly positive somewhere on each leaf, whence $H$
would be space-like there. By requiring that $H$ be space-like
everywhere we are ignoring the case in which portions of
marginally trapped surfaces lie on a space-like horizon and the
remainder on a null horizon. This case will be discussed elsewhere
\cite{ahk} but we will comment on how some of the main results are
modified in this case. Finally, the assumption that $H$ is
space-like also rules out situations in which the horizon reaches
equilibrium and the energy flux across entire cross-sections
vanishes. These will be considered in section \ref{s6}.

To summarize, apart from the possibility that $H$ may be partially
null as discussed separately in  section \ref{s6} and in
\cite{ahk}, for evolving black holes the conditions imposed in
Definition 1 are natural and incorporate most of the physical
situations we have in mind. The world-tubes $\tau_{\rm AH}$ of
apparent horizons resulting from `nice' foliations of numerically
simulated space-times will probably satisfy our conditions and
qualify as dynamical horizons. (For random foliations, the
intuitive condition ${\cal L}_n \Theta_{(\ell)} <0 $ may be
violated, whence $\tau_{\rm AH}$ may well be partially time-like.)
But the notion of dynamical horizons appears to be more general in
the sense that we do not know of a result to the effect that given
a dynamical horizon $H$, the space-time must admit a foliation for
which cross-sections $S$ of $H$ are apparent horizons (rather than
just marginally trapped surfaces, which they certainly are).
Finally, explicit examples of dynamical horizons are provided by
the Vaidya metrics discussed in some detail in Appendix \ref{a1}.
(In this case, the topology of the cross-sections $S$ is $S^2$ and
the generic condition ${\cal L}_{n}\, \Theta_{(\ell)} <0 $ is
satisfied in the dynamical black hole region.) Thus, overall, the
requirements in the Definitions are rather mild. In the remainder
of this paper we will see that the conditions are also
sufficiently strong in the sense that the Definition has a rich
variety of consequences.

\subsection{Hayward's trapping horizons}
\label{s2.2}

To capture the notion of a black hole without reference to
infinity, Hayward \cite{sh} constructed an ingenious quasi-local
framework. Dynamical horizons are closely related to his notion of
trapping horizons. In this subsection, we will clarify the
relation between the two. This discussion will be especially
useful to section \ref{s6} because trapping horizons provide a
natural arena for analyzing the transition at late times from
dynamical to isolated
horizons.\\
\textbf{Definition 2}: A \emph{future, outer, trapping horizon}
(FOTH) is a 3-manifold, $H^\prime$, foliated by closed surfaces
$S^\prime$ such that: i) the expansion of one future directed null
normal $\ell^a$ to the foliation vanishes, $\Theta_{(\ell)} =0$;
ii) the expansion of the other future directed null normal, $n^a$
is negative, $\Theta_{(n)}<0$; iii) the directional derivative of
$\Theta_{(\ell)}$ along $n^a$ is negative; ${\cal L}_{n}\,
\Theta_{(\ell)} <0$.

Here, condition ii) captures the idea that $H^\prime$ is a future
horizon (i.e., of the black hole rather than white hole type) and
condition iii) encodes the idea that it is `outer' and serves to
distinguish black hole type horizons from certain cosmological
ones \cite{sh} which are not ruled out by condition ii).

Our discussion of section \ref{s2.2} shows that $H^\prime$ is
either space-like or null, being  null if and only if the shear
$\sigma$ of $\ell^a$ as well as the matter flux
$T_{ab}\ell^a\ell^b$ across $H$ vanishes. A \emph{space-like} FOTH
is a dynamical horizon on which the additional condition ${\cal
L}_{n}\, \Theta_{(\ell)} <0$ holds. Similarly, a dynamical horizon
\emph{satisfying} ${\cal L}_{n}\, \Theta_{(\ell)} <0$ is a
space-like FOTH. Thus, while neither Definition implies the other,
there is a large overlap between dynamical horizons and FOTHs.  In
generic \emph{dynamic} situations \emph{pertaining to black
holes}, one is likely to encounter horizons which satisfy both
sets of conditions, i.e., lie in the \emph{intersection} of the
two sets. In fact, since one expects the region to the immediate
future of the dynamical horizon to be trapped, a stronger version
of ${\cal L}_n\, \Theta_{(\ell)} <0$ should be satisfied: if
$\th^a$ is a future directed normal to $H$ and $W^a$ is \emph{any}
vector such that $W^a \th_a <0$, then ${\cal L}_W\,
\Theta_{(\ell)}<0$.

The advantage of Definition 1 is that it refers only to the
intrinsic structure of $H$, without any conditions on the
evolution of fields in directions transverse to $H$. As we will
see, this makes it natural to analyze the structure of $H$ using
only the constraint (or initial value) equations. Reciprocally,
Definition 2 has the advantage that it permits $H'$ to be
space-like \emph{or} null. In a spherical collapse of a scalar
field, for example, while $H$ is useful only in the regions where
the flux of the scalar field energy across $H^\prime$ is non-zero,
$H^\prime$ is useful also in the region where it vanishes and the
horizon becomes null. (See section \ref{s6} and the explicit
examples discussed in Appendix \ref{a1}.)

Finally, we recall Hayward's \cite{sh} notions related to a
trapping boundary. A \emph{trapped region} is a connected subset
of space-time through each point $p$ of which there passes a
closed trapped surface (such that $\Theta_{(\ell)} <0$ and
$\Theta_{(n)} <0$). An \emph{inextendable} trapped region
$\textbf{T}$ is a trapped region that cannot be extended. A
\emph{trapping boundary} $\partial \textbf{T}$ is the boundary of
an inextendable trapped region $\textbf{T}$. Physically,
$\textbf{T}$ can be regarded as a black hole region of the
space-time and $\partial \textbf{T}$, as the surface of that black
hole. To establish a desired property of this surface, Hayward had
to introduce a further technical notion: A \emph{limit section} of
the trapping boundary is a smooth, closed sub-manifold of
$\partial \textbf{T}$ which can be obtained as an uniform limit of
closed trapped surfaces lying in $\textbf{T}$. With these
definitions at hand, Hayward showed that if a trapping boundary is
smooth and foliated by limit sections, then the following
conditions hold on each leaf : i) The expansion of one of the null
normal, say $\ell^a$ vanishes; $\Theta_{(\ell)} =0$; ii) The
expansion of the second null normal satisfies $\Theta_{(n)} \le
0$; and iii) ${\cal L}_n\,\Theta_{(\ell)} \le 0$. Thus, if we
ignore the degenerate cases where equalities hold in the last two
equations, the boundary is a FOTH. In this sense then,
generically, if the black hole is genuinely dynamical, the
trapping boundary $\partial \textbf{T}$ would be a dynamical
horizon, and if it has reached equilibrium, it would be a weakly
isolated horizon \cite{afk}. In the former case, $\partial
\textbf{T}$ would represent the canonical dynamical horizon
associated with the black hole under consideration.

\subsection{Notation and strategy}
\label{s2.3}

In the next four sections of this paper, we will consider a
dynamical horizon $H$ and explore its properties. If $H$ admits
more than one foliation by marginally trapped surfaces satisfying
Definition 1, we will just choose any one of them and use it
throughout our calculations. Our results will apply to all such
foliations. At appropriate places, we will comment on the
expressions which are foliation independent.  Leaves of the fixed
foliation will be called \emph{cross-sections} of $H$.

Let us begin by specifying notation. For simplicity, All manifolds
will be assumed to be smooth (i.e. $C^{k+1}$ with $k\ge 3$) and
orientable and all fields will be assumed to be smooth (i.e.,
$C^k$). The space-time metric $g_{ab}$ has signature $(-,+,+,+)$
and its derivative operator will be denoted by $\nabla$. The
Riemann tensor is defined by $R_{abc}{}^d W_d := 2 \nabla_{[a}
\nabla_{b]}W_c$, the Ricci tensor by $R_{ab} := R_{acb}{}^c$ and
the scalar curvature by $R := g^{ab} R_{ab}$. We will assume the
field equations
\be \label{fe} R_{ab}- \frac{1}{2}R\, g_{ab} + \Lambda g_{ab} =
8\pi G T_{ab}\, .\ee
(With these conventions, de Sitter space-time has positive
cosmological constant $\Lambda$.) We assume that $T_{ab}$
satisfies the dominant energy condition (although, as the reader
can easily tell, several of the results will hold under weaker
restrictions.) To keep the discussion reasonably focussed, we will
not consider gauge fields with non-zero charges on the horizon.
Inclusion of these fields is not difficult but introduces a number
of subtleties and complications which are irrelevant for numerical
relativity and astrophysics. They will be discussed elsewhere.

\begin{figure}
  \begin{center}
  \includegraphics[height=12cm]{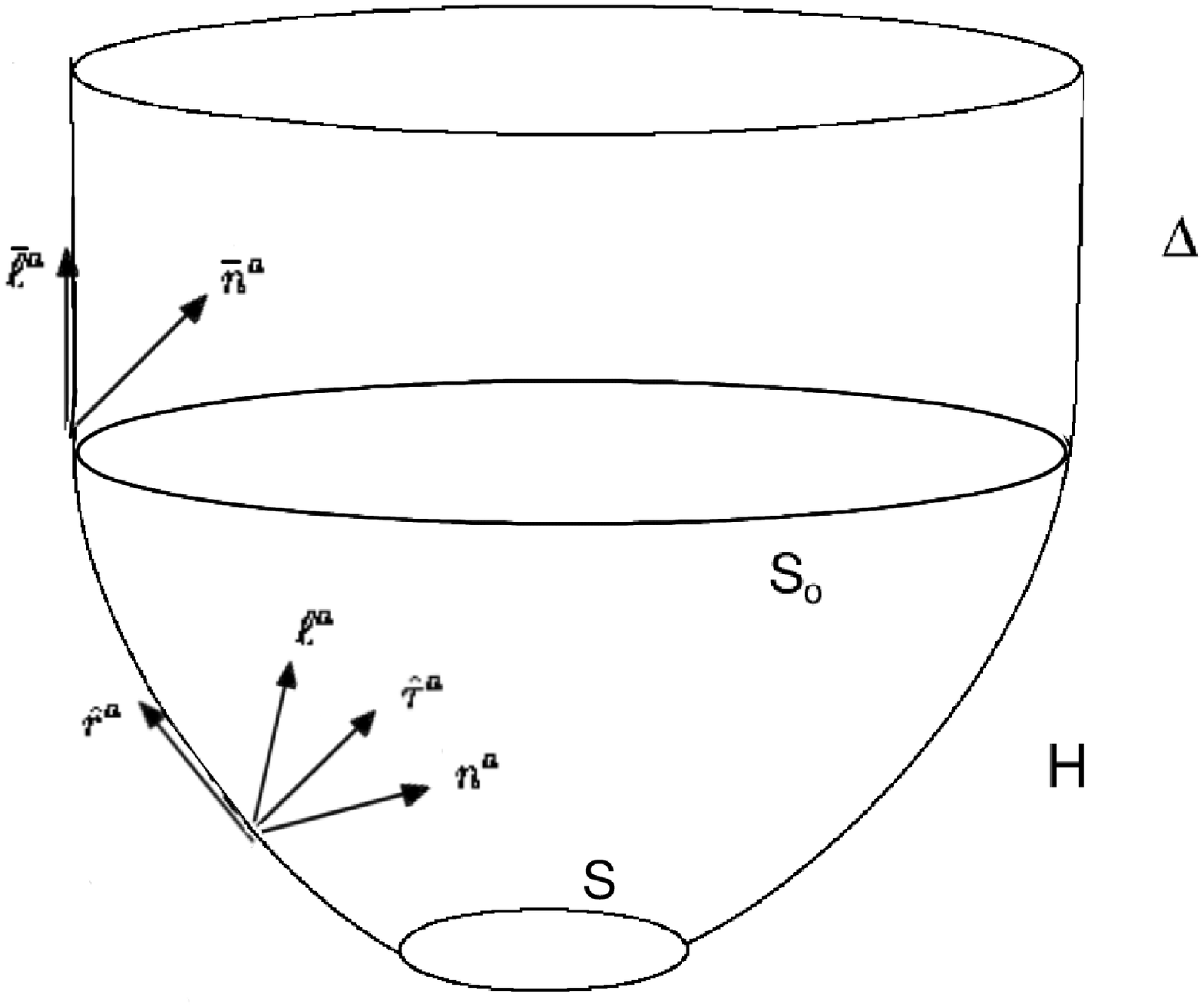}
  \caption{$H$ is a dynamical horizon, foliated by marginally
trapped surfaces $S$. $\th^a$ is the unit time-like normal to $H$
and $\rh^{\,a}$ the unit space-like normal within $H$ to the
foliations. Although $H$ is space-like, motions along $\rh^{\,a}$
can be regarded as time evolution with respect to observers at
infinity. In this respect, one can think of $H$ as a hyperboloid
in Minkowski space and $S$ as the intersection of the hyperboloid
with space-like planes. $H$ joins on to a weakly isolated horizon
$\Delta$ with null normal $\bar{\ell}^a$, at a cross-section $S_o$.}\label{dhfig}
  \end{center}
\end{figure}
Geometry of the dynamical horizon $H$ is pictorially represented
in figure \ref{dhfig}. The unit normal to $H$ will be denoted by $\th^a$;
$g_{ab}\th^a\th^b = -1$. The intrinsic metric and the extrinsic
curvature of $H$ are denoted by $q_{ab}:= g_{ab} + \th_a\th_b$ and
$K_{ab}:={q_a}^c{q_b}^d\grad_c\th_d$ respectively. $D$ is the
derivative operator on $H$ compatible with $q_{ab}$,  $\R_{ab}$
its Ricci tensor and $\R$ its scalar curvature.  The unit
space-like vector orthogonal to $S$ and tangent to $H$ is denoted
by $\rh^{\,a}$. Quantities intrinsic to $S$ will be generally
written with a tilde. Thus, the two-metric on $S$ is $\q_{ab}$ and
the extrinsic curvature of $S\subset H$ is
$\K_{ab}:=\widetilde{q}_a^{\,\,\,\,c}\widetilde{q}_b^{\,\,\,\,d}
D_c\rh_d$; the derivative operator on $(S, \q_{ab})$ is
$\widetilde{D}$ and its Ricci tensor is $\twoR_{ab}$. Finally, in
the next four sections we will fix the rescaling freedom in the
choice of null normals via $\l^a:=\th^{\,a}+\rh^{\,a}$ and
$n^a:=\th^{\,a}-\rh^{\,a}$ (so that $\ell^a n_a = -2)$.  This
convention will have to be modified in the discussion of
transition to equilibrium of section \ref{s6}.

We first note an immediate consequence of the definition. Since
$\Theta_\ls =0$ and $\Theta_\ns <0$, it follows that
\be \K = \tilde{q}^{ab} D_a \rh_b = \frac{1}{2}\, \tilde{q}^{ab}
\nabla_a (\ell_b -n_b) = - \frac{1}{2} \Theta_{(n)} >0. \ee
Hence the area $a_S$ of $S$ increases monotonically along $\rh^{\,
a}$. Thus the second law of black hole mechanics holds on $H$. Our
first task is to obtain an explicit expression for the change of
area.

Our main analysis is based on the fact that, since $H$ is a
space-like surface, the Cauchy data $(q_{ab},K_{ab})$ on $H$ must
satisfy the usual scalar and vector constraints
\begin{eqnarray} H_S &:=& \R + K^2 - K^{ab}K_{ab}
          = 16\pi G \T_{ab}\,\th^{\,a}\th^{\,b}
            \label{hamconstr}\\
    H_V^a &:=& D_b\left(K^{ab} - Kq^{ab}\right)
    = 8\pi G \T^{bc}\,\th_{\, c}{q^a}_b
    \label{momconstr} \, .\end{eqnarray}
where
\be \label{T} \T_{ab} = T_{ab} - \frac{1}{8\pi G}\, \Lambda g_{ab}
\ee
and $T_{ab}$ is the matter stress-energy tensor. The strategy
behind the key calculations in the next three sections is entirely
straightforward:  We will fix two cross-sections $S_1$ and $S_2$
of $H$, multiply $H_S$ and $H_V^a$ with appropriate lapse and
shift fields and integrate the result on a portion $\Delta H
\subset H$ which is bounded by $S_1$ and $S_2$.

\emph{Remark}: As noted in section \ref{s2.2}, the notions of
dynamical horizons and FOTHs are closely related and, in
physically interesting situations involving evolving black holes,
both sets of conditions will be satisfied. However, there are key
differences between our analysis based on dynamical horizons and
Hayward's analysis \cite{sh} based on FOTHs . While our analysis
will be based on the standard 3+1 decomposition, Hayward's
framework is based on a 2+2 decomposition. The 2+2 framework is
better suited for analyzing more general horizons where $H$ is
partially time-like and partially null but has the disadvantage
that it fails to make it manifestly clear that the fields of
interest are defined just by the horizon geometry and are
independent of extensions used off $H$. In terms of results, our
final result on the topology of cross-sections is the same as that
of \cite{sh}. However, results in the rest of the paper are quite
different. Specifically, our flux formulae are new, our discussion
includes angular momentum, our generalization of black hole
mechanics is different, and our definition of the horizon energy
and balance laws are new.

\section{Energy fluxes and area balance}
\label{s3}

Let us now turn to the task of relating the change in area to the
flux of energy across $H$. Along the way, we will establish that
the topology of the cross sections $S$ of $H$ is severely
restricted in the case when $\Lambda \ge 0$.

\subsection{Area increase and topology of $S$}
\label{s3.1}

As is usual in general relativity, the notion of energy is tied to
a choice of a vector field. The definition of a dynamical horizon
provides a preferred direction field; that along $\ell^a$. To fix
the proportionality factor, or the lapse $N$, let us first
introduce the area radius $R$, a function  which is constant on
each $S$ and satisfies $a_S = 4\pi R^2$. Since we already know
that area is monotonically increasing, $R$ is a good coordinate on
$H$. Now, the 3-volume $d^3V$ on $H$ can be decomposed as $d^3V =
|\partial R|^{-1}dR d^2V$ where $\partial$ denotes the gradient on
$H$.  Therefore, as we will see, our calculations will simplify if
we choose $N_R = |\partial R|$. In this sub-section, we will make
this simple choice, obtain an expression for the change in area
and show that the topology of the cross-sections $S$ is severely
restricted. In section \ref{s3.3} we will generalize this \emph{area
balance law} to include a more general family of lapses.

Since the area increase formula plays an important role throughout
the paper, we will provide a detailed derivation. Fix two cross
sections $S_1$ and $S_2$ of $H$ and denote by $\Delta H$ the
portion of $H$ they bound. We are interested in calculating the
flux of energy associated with $\xi_{(R)}^a = N_R \ell^a$ across
$\Delta H$. Denote the flux of \emph{matter} energy across $\Delta
H$ by $\mathcal{F}^{(R)}_\m$:
\be \mathcal{F}^{(R)}_\m := \int_{\Delta H}
T_{ab}\th^{\,a}\xi_{(R)}^b d^3V.\ee
By taking the appropriate combination of (\ref{hamconstr}) and
(\ref{momconstr}) we obtain
\ba \label{eq:fluxTr} \mathcal{F}^{(R)}_\m &=& \frac{1}{16\pi G}
\int_{\Delta H}\, N_R \left(H_S + 2\rh_a H_V^a \right)\, d^3V
\nonumber\\
&=&\frac{1}{16\pi G} \int_{\Delta H}\, N_R\left(\R + K^2 -
K^{ab}K_{ab} + 2\rh_aD_bP^{ab}\right) \, d^3V \ea
where $P^{ab}$ is defined as
\be P^{ab} = K^{ab} - Kq^{ab} \, . \ee
Since $H$ is foliated by compact 2-manifolds $S$, we can perform a
$2+1$ decomposition of various quantities on $H$.  First, the
Gauss-Codacci equation relating the space-time curvature to the
intrinsic curvature of $H$ leads to
\be \label{eq:gausscodacci} 2\G_{ab}\rh^a\rh^b = -\twoR + \K^2 -
\K_{ab}\K^{ab} \ee
where $\G_{ab}$ is the Einstein tensor of $(H,q_{ab})$, and the
definition of the Riemann tensor gives
\be \label{eq:riemann} \R_{ab}\rh^a\rh^b =
-2\rh^aD_{[a}D_{b]}\rh^b = D_a\alpha^a + \K^2 - \K_{ab}\K^{ab} \ee
where
\be \label{eq:alpha} \alpha^a := \rh^bD_b\rh^a - \rh^aD_b\rh^b\, .
\ee
Combining equations (\ref{eq:gausscodacci}) and
(\ref{eq:riemann}), we can obtain a useful expression relating the
scalar curvatures on $H$ and $S$:
\be \label{eq:threeR} \R = 2(\R_{ab}-\G_{ab})\rh^a\rh^b = \twoR +
\K^2 -\K_{ab}\K^{ab} +2D_a\alpha^a\, . \ee
Transvecting the momentum constraint equation with $\rh_b$ gives
\be \label{eq:rbmomconstr} \rh_bD_aP^{ab} = D_a\beta^a -
P^{ab}D_a\rh_b \ee
where
\be \label{eq:beta} \beta^a := K^{ab}\rh_b - K\rh^a \, . \ee
Substituting the results of equations (\ref{eq:threeR}) and
(\ref{eq:rbmomconstr}) into the integrand of the right side of eq.
(\ref{eq:fluxTr}) yields
\be \label{general}  H_S+ 2\rh_a H^a_V = \twoR + \K^2 -
\K_{ab}\K^{ab} + K^2 - K_{ab}K^{ab} - 2P^{ab}D_a\rh_b +
2D_a\gamma^a  \ee
where
\be \label{eq:gamma} \gamma^a := \alpha^a+\beta^a \, . \ee
For further simplification, let us bear in mind that we will
eventually use the key property that the cross sections $S$ are marginally
trapped surfaces, i.~e. $\Theta_\ls = 0$. In terms of the
extrinsic curvatures $K_{ab}$ and $\K_{ab}$, the expansion can be
written as
\be \label{eq:expansion0} \Theta_{(\ell)} = K -
K_{ab}\rh^{\,a}\rh^{\,b} +\K  \,. \ee
To recast the extrinsic curvature terms in Eq. (\ref{general})
using $\Theta_{(\ell)}$, it is convenient to perform a
decomposition of the two extrinsic curvatures:
\ba
\K_{ab} &=& \frac{1}{2}\K\q_{ab} + \S_{ab} \\
K_{ab} &=& A\q_{ab} + S_{ab} + 2\W_{(a}\rh_{b)} + B\rh_a\rh_b \, ,
\ea
where $\S_{ab}$ is the trace-free part of $\K_{ab}$; $S_{ab}$, the
trace-free part of the projection of $K_{ab}$ into $S$; $\W_{a}$
is the projection of $K_{ab}\rh^b$ into $S$; $A := \frac{1}{2}
K_{ab}\q^{ab}$ and $B :=K_{ab}\rh^a\rh^b$. Note that $S_{ab}$,
$\S_{ab}$ and $\W_{a}$ are \emph{two-dimensional} tensors
intrinsic to the cross-section $S$. Substituting the above
decompositions in eq. (\ref{general}) $\mathcal{F}^{(R)}_\m$ and
using eq. (\ref{eq:expansion0}), we obtain
\ba \label{general2}H_S+ 2\rh_a H^a_V &=&  \twoR -
\sigma_{ab}\sigma^{ab} - 2\W_a\W^a - 2\W^a\rh^bD_b\rh_a \nonumber
\\ && +  \frac{1}{2}\, \Theta_{(\ell)}\, (\Theta_{(\ell)}
+4B) +2D_a\gamma^a  \ea
where $\sigma_{ab} = S_{ab} + \S_{ab}$ is the shear of the null
vector $\ell^a = \th^a+\rh^a$;\, i.e. $\sigma_{ab} := \q_a{}^m
\q_b{}^n \nabla_m \ell_n - \frac{1}{2}\, \q_{ab} \q^{mn} \nabla_m
\ell_n$. Our task in the remainder of this calculation is to
simplify the right side of this equation.

With this goal in mind, let us now turn our attention to the
vector $\gamma^a$ defined in eqs. (\ref{eq:gamma}),
(\ref{eq:alpha}), and (\ref{eq:beta}):
\ba \label{eq:projgamma} \gamma^a = \alpha^a + \beta^a
&=&\rh^bD_b\rh^a - \rh^aD_b\rh^b + K^{ab}\rh_b - K\rh_a \nonumber\\
&=& \rh^bD_b\rh^a + \W^a - \Theta_{(\ell)} \rh^a\, . \ea
Finally, it is convenient to re-express the acceleration term as
\be \label{eq:dnr} \rh^aD_a\rh_b = (N_R)^{-1}\, \twoD_b N_R \, .
\ee
Then, Eq. (\ref{general2}) can be rewritten as:
\ba \label{general3} H_S+ 2\rh_a H^a_V &=& \twoR -
\sigma_{ab}\sigma^{ab} - 2 \zeta^a \zeta_a + 2 \twoD_a \zeta^a
\nonumber\\
&+& \frac{1}{2}\Theta_{(\ell)} (4K-3\Theta_{\ell)}) - 2 \rh^a D_a
\Theta_{(\ell)}
\ea
where the vector $\zeta^a$, tangent to the cross sections, is
defined as
\be \zeta^a := \W^a + \twoD^a\ln N_R =
\widetilde{q}^{\,ab}\rh^{\,c}\grad_c\l_b\, . \ee
Equation (\ref{general3}) is completely general; it holds on any foliated
space-like surface. We now wish to use the fact that surface of
interest is in fact a dynamical horizon. Integrating on the
portion $\Delta H$ of the horizon $H$, using the fact that the
cross-sections $S$ are compact and $\Theta_{(\ell)}$ vanishes, we
are led to a remarkably simple result:
\be \label{eq:simpleflux} \mathcal{F}^{(R)}_\m = \frac{1}{16\pi
G}\int_{\Delta H} N_R\left( \twoR - \sigma_{ab}\sigma^{ab} - 2
\zeta^a \zeta_a \right) \, d^3V  \ee
%
Using the abbreviations $|\sigma|^2 := \sigma_{ab}\sigma^{ab}$ and
$|\zeta|^2 := \zeta_a\zeta^a$, this  can be
rewritten as
\be \label{eq:NR} \int_{\Delta H} N_R \twoR\,d^3V = 16\pi G
\int_{\Delta H} \T_{ab}\th^{\,a}\xi_{(R)}^b\,d^3V  + \int_{\Delta
H} N_R\left\{ |\sigma|^2 + 2|\zeta|^2\right\}\,d^3V \, .\ee
This is the key equation we were seeking to obtain quantitative
expression for the change in the horizon area in fully dynamic
processes. It will have several important applications. In the
remainder of this sub-section we will focus on the first of these:
its implications for the topology of $S$.

Let us first recall that the volume element $d^3V$ on $H$ can be
written as $d^3V = N_R^{-1}dR\,d^2V$ where $d^2V$ is the area
element on $S$. Therefore, the integral on the left hand side
becomes:
\be \int_{\Delta H} N_R \twoR\,d^3V = \int_{R_1}^{R_2} dR\, \oint
\twoR\, d^2V = \mathcal{I} (R_2 - R_1)\, .\ee
Here $R_1$ and $R_2$ are the (geometrical) radii of $S_1$ and
$S_2$; we have used  the Gauss-Bonnet theorem in the second step;
and, $\mathcal{I}$ is the Gauss invariant of the closed,
orientable 2-manifold $S$. (Our choice of lapse was made to enable
this step in the calculation.) Substituting back in eq.
(\ref{eq:NR}) we obtain:
\be\label{ab1}  \mathcal{I}\, (R_2 - R_1) =  16\pi G \int_{\Delta
H} (T_{ab}- \frac{\Lambda}{8\pi G} g_{ab})
\th^{\,a}\xi_{(R)}^b\,d^3V + \int_{\Delta H} N_R\left\{ |\sigma|^2
+ 2|\zeta|^2\right\}\,d^3V \ee
where we have used the definition (\ref{T}) of $\T_{ab}$. The
discussion of topology of $S$ is naturally divided in to three
cases, depending on (the sign of) the cosmological constant.
\begin{description}
\item[\emph{Case 1}:] $\Lambda > 0$. Now, since the stress energy
tensor $T_{ab}$ is assumed to satisfy the dominant energy
condition, the right side is manifestly positive definite. Since
we already know that area increases along $\rh^a$, we have $R_2 -
R_1
>0$. Hence it follows that $\mathcal{I}$ must be positive,
whence the closed, orientable 2-manifolds $S$ are necessarily
topological 2-spheres and $\mathcal{I} = 8\pi$. Eq. (\ref{ab1})
now becomes:
\ba\label{ab2}  \frac{R_2-R_1}{2G} &=&  \int_{\Delta H}
(T_{ab}- \frac{\Lambda}{8\pi G}\, g_{ab})
\th^{\,a}\xi_{(R)}^b\,d^3V \nonumber \\ &+& \frac{1}{16\pi G}\,\int_{\Delta H}
N_R\left\{ |\sigma|^2 + 2|\zeta|^2\right\}\,d^3V\,.
\ea
\item[\emph{Case 2}]: $\Lambda =0$. Now the right side of eq. (\ref{ab1})
is necessarily non-negative. Hence, the topology of $S$ is either
that of a 2-sphere (if the right side is positive) or that of a
2-torus (if the right side vanishes). As mentioned in Section
\ref{s2.3}, this constraint on topology was obtained by Hayward
\cite{sh} using a 2+2 framework.

The torus topology can occur if and only if $T_{ab}\ell^b$,
$\sigma_{ab}$ and $\zeta^a$ all vanish everywhere on $H$. Going
back to Eq. (\ref{general2}), we conclude that the scalar
curvature $\twoR$ of $S$ must also vanish
on every cross-section.%
\footnote{We thank J. Lewandowski for this observation. In view of
these highly restrictive conditions, toroidal dynamical horizons
appear to be unrelated to the toroidal topology of cross-sections
of the event horizon discussed by Shapiro, Teukolsky, Winicour and
others \cite{st,jw}.}
Also, using the fact that $H$ is space-like, it now follows from
eq. (\ref{spacelike}) that in this case ${\cal L}_n\,
\Theta_{(\ell)} =0$ everywhere on $H$. Thus, in this case the
dynamical horizon cannot be a FOTH. Furthermore, since
$\Theta_{(\ell)}, \sigma_{ab}$ and $R_{ab}\ell^b$ all vanish on
$H$, the Raychaudhuri equation now implies that ${\cal L}_\ell\,
\Theta_{(\ell)}$ also vanishes. These strong restrictions imply
that this is a degenerate case. For such horizons, although we
know that the area must increase, eq. (\ref{ab1}) trivializes
whence we do not have a quantitative formula for the amount by
which the area increases.

For generic dynamical horizons, the topology is $S^2$ and the
quantitative relation is given by:
\be\label{ab3}  \frac{1}{2G} (R_2 - R_1) =  \int_{\Delta H} T_{ab}
\th^{\,a}\xi_{(R)}^b\,d^3V + \frac{1}{16\pi G}\,\int_{\Delta H}
N_R\left\{ |\sigma|^2 + 2|\zeta|^2\right\}\,d^3V \, .\ee

\item[\emph{Case 3}:] $\Lambda <0$. In this case there is no control on
the sign of the right hand side of eq. (\ref{ab1}). Hence, a
priori any topology is permissible. Stationary solutions with
quite general topologies are known for black holes which are
locally asymptotically anti-de Sitter. Event horizons of these
solutions are the potential asymptotic states of these dynamical
horizons in the distant future.
\end{description}
\emph{In the remainder of this paper we will restrict our detailed
calculations to the case of 2-sphere topology.}

\textsl{Remark:} The above considerations provide an interesting
constraint on the topology of marginally trapped surfaces if
$\Lambda \ge 0$. As it stands, the discussion is restricted to the
topology of cross-sections of dynamical horizons $H$. However, it
is straightforward to generalize these results. Consider
\emph{any} 3-manifold $\bar H$, foliated by compact 2-surfaces
$\bar{S}$. Then, by integrating (\ref{general3}) only on one leaf
$\bar{S}$ of the foliation (rather than on $\Delta{\bar{H}}$), in
place of (\ref{ab1}) we obtain:
\ba\label{ab4}  \mathcal{I}\,  &=&  16\pi G \int_{\bar{S}} (T_{ab}
- \frac{\Lambda}{8\pi G} g_{ab}) \th^{\,a}\ell^b\,d^2V
\nonumber\\
&+& \int_{\bar{S}} \left( |\sigma|^2 + 2|\zeta|^2 -
\frac{1}{2}\,\Theta_{(\ell)}\,( 4K-3\Theta_{(\ell)} ) + 2\rh^a D_a
\Theta_{(\ell)} \right)\,d^2V \, .\ea
Now, if one leaf $\bar{S}_o$ of the foliation is marginally
trapped and if $\rh^a D_a \Theta_{(\ell)} \ge 0$ on $\bar{S}_o$,
we conclude that the topology of $\bar{S}_o$ must be that of a
$S^2$ if $\Lambda >0$ and of a $S^2$ or a $T^2$ if $\Lambda =0$.
$T^2$ is a degenerate case in the sense explained above.  Note
that no assumption on the expansion $\Theta_{(n)}$ of $n^a$ has
been used here.

Since $\bar{H}$ was arbitrary, we can also reach a conclusion on
the topology of any marginally trapped surface $S$ in a space-time
satisfying the dominant energy condition: either $S$ is
topologically $S^2$ or $T^2$ or its (first order) deformation
along any space-like, outward direction leads to a trapped
surface. (A space-like direction $V^a$ will be said to be \emph{outward}
if $V^a \ell_a
>0$.) In particular, then, if the topology is more complicated, the
surface cannot lie on a trapping boundary. This is essentially
Hawking's result \cite{swh}.

\subsection{Gravitational energy flux}
\label{s3.2}

Let us now interpret the various terms appearing in the area
balance law. For simplicity of presentation, we will first focus
on the case $\Lambda =0$ and comment on the $\Lambda \not= 0$
cases at the end.

The left side of eq. (\ref{ab3}) provides us with the change in
the horizon radius caused by the dynamical process under
consideration. Since the expansion $\Theta_{(\ell)}$ vanishes,
this is also the change in the \emph{Hawking mass} as one moves
from the cross section $S_1$ to $S_2$. The first integral on the
right side of this equation is the flux $\mathcal{F}^{(R)}_\m$ of
matter energy associated with the vector field $\xi_{(R)}^a$. The
second term is purely geometrical and since it accompanies the
term representing the matter energy flux, we propose to interpret
it as the flux $\mathcal{F}^{(R)}_\g$ of $\xi_{(R)}^a$-energy in
the gravitational radiation:
\begin{equation} \label{gravflux1}
 \mathcal{F}^{(R)}_\g := \frac{1}{16\pi G}
 \int_{\Delta H} N_R\left\{ |\sigma|^2 + 2|\zeta|^2\right\}\,d^3V\, .
\end{equation}
While the interpretation is naturally suggested by the area
balance law (\ref{ab3}), the key question is: Is this proposal
physically viable? The purpose this sub-section is to argue that
the answer is in the affirmative in the sense that it passes the
`standard' tests one uses to demonstrate the viability of the
Bondi flux formula at null infinity.

\b\,\,  \emph{Gauge invariance:} Since we did not have to introduce
any structure,
such as coordinates or tetrads, which is auxiliary to the problem,
the expression is obviously gauge invariant. This is to be
contrasted with definitions involving pseudo-tensors or background
fields.

\b\,\,  \emph{Positivity:} The energy flux is manifestly
non-negative. In the case of the Bondi flux, positivity played a
key role in the early development of the gravitational radiation
theory.  It was perhaps the most convincing evidence that
gravitational waves are not coordinate artifacts but carry
\emph{physical} energy; as Bondi put it, `one can heat water with
them'.

It is surprising that a simple, manifestly non-negative expression
can exist in the strong field regime of dynamical horizons. We did
argue in section \ref{s1} that, since the energy is lost from the
asymptotic region, one does expect an appropriately defined notion
of gravitational energy flux across the surface of the black hole
to be well-defined and positive. But the way in which the details
work out is quite subtle. For example, since the issue is that of
controlling signs, one may be tempted to conjecture that this
positivity is a property of the black hole region where the
expansion $\Theta_{(\ell)}$ of the outgoing normal is
non-positive, i.e., of a definite sign. However, this conjecture
turns out to be false! To show this, let us carry out the analysis
of section \ref{s3.1} on a general, foliated space-like surface
$\bar{H}$. We can still obtain eq. (\ref{general3}) but, as is
clear from (\ref{ab4}), in place of the $\mathcal{F}^{(R)}_\g$ of
(\ref{gravflux1}) the final expression would be:
\begin{equation} \label{gravflux2}
 \bar{\mathcal{F}}^{(R)}_\g := \frac{1}{16\pi G} \int_{\Delta H}
 N_R\left\{ |\sigma|^2 + 2|\zeta|^2  +
 \frac{1}{2}\Theta_{(\ell)} (4K-3\Theta_{(\ell)})
 + 2\rh^a D_a \Theta_{(\ell)}\right\} \,d^3V\, . \end{equation}
The key point is that if $\bar{H}$ is not a dynamical horizon, the
sign of the last two terms cannot be controlled, not even when
$\bar{H}$ lies in the black hole region and is foliated by trapped
(rather than marginally trapped) surfaces $\bar{S}$. Thus, the
positivity of $\mathcal{F}^{(R)}_\g $ is a rather subtle property,
not shared by 3-surfaces which are foliated by non-trapped
surfaces, nor those which are foliated by trapped surfaces; one
needs a foliation \emph{precisely by marginally trapped surfaces}.
Thus, the property is delicately matched to the definition of
dynamical horizons. This is but one instance of the mysterious
ability of Einstein's equations to realize physical expectations
through geometrical structures in completely unforeseen and subtle
ways.%
\footnote{Some of the well-known examples are: the well-posedness
of the Cauchy problem; the positive energy theorems at spatial and
null infinity; positivity of the Bondi flux at null infinity; and
more open-ended issues such as cosmic censorship and Penrose
inequalities. Not only did the list of considerations that led
Einstein to his field equations not include these issues but even
the physical relevance of  most of them was not appreciated for
decades after the discovery of general relativity. Yet, quite
mysteriously, the field equations incorporate them correctly!}

\b\,\, \emph{Locality:} All fields used in it are defined by the
\emph{local} geometrical structures on cross-sections of $H$. This
is a non-trivial property, shared also by the Bondi-flux formula.
However, it is not shared in other contexts. For example, the
proof of the positive energy theorem by Witten \cite{ew} provides
a positive definite energy density on Cauchy surfaces. But since
it is obtained by solving an elliptic equation with appropriate
boundary conditions at infinity, this energy density is a highly
non-local function of geometry. Locality of $\mathcal{F}^{(R)}_\g$
enables to associate it with the energy of gravitational waves
instantaneously falling across any cross section $S$.

\b\,\,  \emph{Vanishing in spherical symmetry:} The fourth
criterion is that the flux should vanish in presence of spherical
symmetry. Suppose the cross-sections $S$ of $H$ are spherically
symmetric. Since the only spherically symmetric vector field and
trace-free, second rank tensor field on a 2-sphere are the zero
fields, $\sigma_{ab}=0$ and $\zeta^a=0$.

\b\,\,  \emph{Relation to perturbation theory:} The fifth criterion
comes from perturbation theory. One
can envisage a situation in which the dynamical horizon is, in an
appropriate physical sense, \emph{weakly} dynamical. In this case, it
can be regarded as a perturbation of a non-expanding horizon
\cite{afk} (see section \ref{s6}). It is then natural to ask if in
this case the gravitational flux (\ref{gravflux1}) reduces to the
expression derived from perturbation theory off Kerr horizons. The
answer is in the affirmative.

\b\,\,  \emph{Balance law:} The Bondi-Sachs energy flux also has
the important property that there is a \emph{locally} defined
notion of the Bondi-energy $E(C)$ associated with any 2-sphere
cross-section $C$ of future null infinity and the difference
$E(C_1) - E(C_2)$ equals the Bondi-Sachs flux through the portion
of null infinity bounded by $C_2$ and $C_1$. Does the expression
(\ref{gravflux1}) share this property? The answer is in the
affirmative: as noted in the beginning of this sub-section, the
integrated flux is precisely the difference between the
\emph{locally defined} Hawking mass associated with the
cross-section. In Section \ref{s5} we will extend these
considerations to include angular momentum.

\b\,\,  \emph{Hamiltonian interpretation:} Finally, the
Bondi-Sachs energy flux has an additional attractive property
which supports its interpretation, although it is not  a direct,
physical, viability criterion: Using a Hamiltonian framework, one
can show that it is the generator of a Bondi-Metzner-Sachs
time-translation on the gravitational phase space \cite{as,wz}.
Does the gravitational flux (\ref{gravflux1}) also enjoy this
property? Recently, Booth and Fairhurst \cite{bf2} have shown that
the answer is in the affirmative.

It is very surprising that there should be a meaningful expression
for the gravitational energy flux in the strong field regime where
gravitational waves can no longer be envisaged as ripples on a
flat space-time. Taken together, the properties discussed above
provide a strong support in favor of the interpretation of
(\ref{gravflux1}) as the $\xi_{(R)}$-energy flux of carried by
gravitational waves into the portion $\Delta H$ of the dynamical
horizon. Nonetheless, it is important to continue to think of new
criteria and make sure that (\ref{gravflux1}) passes these tests.
For instance, in physically reasonable, stationary, vacuum
solutions to Einstein's equations, one would expect that the flux
should vanish. However, on dynamical horizons the area must
increase. Thus, one is led to conjecture that these space-times do
not admit dynamical horizons. While special cases of this
conjecture have been proved, a general proof is still lacking.

So far, we have set the cosmological constant $\Lambda$ to zero.
Even when $\Lambda$ is non-zero, it seems natural to continue to
interpret (\ref{gravflux1}) as the $\xi_{(R)}$-energy flux of
carried by gravitational waves into the portion $\Delta H$ of the
dynamical horizon. However, now there is an additional, purely
geometrical contribution to the area change of eq. (\ref{ab2})
coming from the cosmological repulsion or attraction induced by
the cosmological constant. If $\Lambda$ is positive, the area of
the cross-sections $S$ of $H$ would continue to grow just because
of the cosmological expansion even when there is no flux of
gravitational or mater energy across $\Delta H$, while if
$\Lambda$ is negative, it would decrease.

To conclude this sub-section, we will comment on some issues
related to the physical interpretation of the flux formula. Note
first that the flux refers to a \emph{specific} vector field
$\xi^a_{(R)}$ and measures the change in the Hawking mass
associated with the cross-sections. This need not be a good
measure of the physical mass in presence of angular momentum
(see section \ref{s5}). Secondly, one can envisage a situation in
which the portion $\Delta H$ bounded by $S_2$ and $S_1$ of a
dynamical horizon admits two \emph{distinct} foliations in the
both of which share the leaves $S_1$ and $S_2$, or, a situation in
which two distinct dynamical horizons $H_1$ and $H_2$ share the
2-spheres $S_2$ and $S_1$. In these cases, the observer fields
$\xi_{(R)}$ are distinct. Although the \emph{total} fluxes
corresponding to the two fields do agree  ---they are given by the
change in horizon radius as one goes from $S_1$ to $S_2$--- the
split between the matter contribution and the gravitational wave
contribution would be different. This is not surprising because we
are in a strong field region and it is not inappropriate for two
observers to disagree on how much energy is contained in matter
and how much in gravitational radiation. Indeed, a priori, what is
surprising is that the sum of the two contributions is the same,
i.e., there is an area balance law. Nonetheless, while
interpreting fluxes, the fact that the energy refers to specific
observers defined on $H$ is an important caveat that should be
kept in mind.

Next, let us consider the various terms in the integrand of our
flux formula (\ref{gravflux1}). The presence of the shear term
$|\sigma|^2$ seems natural from one's expectations based on
perturbation theory at the event horizon of the Kerr family
\cite{hh,chandra}. What about the term $|\zeta|^2$? Since
$\zeta^a = \q^{an}\rh^m\nabla_m \ell_n$, this term could arise
only because $H$ is space-like rather than null: On a null
surface, the analog of $\rh^a$ is parallel to $\ell^a$, whence the
analog of $\zeta^a$ vanishes identically.  To bring out this
point, let us consider a more general case than the one considered
in this paper and allow the cross-sections $S$ to lie on a horizon
which is partially null and partially space-like. Then, using a
2+2 formulation used by Hayward, one can conclude that flux on the
null portion is given \emph{entirely} by the term $|\sigma|^2$
\cite{ahk}. However, on the space-like portion, the term
$|\zeta|^2$ does not in general vanish. Indeed, on a dynamical
horizon, it \emph{cannot} vanish in presence of rotation: the
angular momentum is given by the integral of $\zeta^a \varphi_a$,
where $\varphi^a$ is the rotational symmetry.

\subsection{Generalization of the area balance law}
\label{s3.3}

At future null infinity $\mathbb{I}^+$, there is a well-defined,
4-dimensional translation sub-group $\mathbb{T}$ of the asymptotic
symmetry group (called the Bondi-Metzner-Sachs group) and there is
a well-defined notion of energy associated with each time
translation in $\mathbb{T}$. Observers following these vector
fields can be physically interpreted as the asymptotically
inertial ones. In sections \ref{s3.1} and \ref{s3.2}, we
associated energy with observers following the vector fields $N_R
\ell^a$. Are there more general families with which we can
similarly assign a notion of energy?

At the dynamical horizon $H$ we are in the strong field regime,
whence there is no longer a universal group of horizon symmetries.
But we can build intuition from the well-developed theory of
weakly-isolated horizons $\Delta$. In this case, to begin with,
one encounters three universality classes of horizon symmetries
\cite{abl2}. Physically, the most interesting case is that of type
II isolated horizons in which the symmetry group is 2-dimensional,
with generators $c \ell^a + \Omega \fie^a$, where $c, \Omega$ are
constants, while $\ell^a, \fie^a$ are tangential to $\Delta$ and
generate a combination of a time translation and a rotation. In
globally stationary, axi-symmetric space-times, these are
restrictions to $\Delta$ of the two Killing fields but generically
they are defined just at the horizon. Nonetheless, they can be
used very effectively in the Hamiltonian framework to introduce
the notion of the horizon energy and angular momentum. For
dynamical horizons $H$, it is natural to extend these notions in
such a way that when $H$ reaches equilibrium and becomes an
isolated horizon, the dynamical horizon framework tends to the
isolated horizon one. An obvious strategy is to make the
coefficients $c$ and $\Omega$ dynamical, i.e., $R$-dependent. In
this sub-section we will focus only on the analog of the
coefficient $c$, i.e., ignore rotation as in \cite{afk}. Inclusion
of rotation and the analog of $\Omega$ will be carried out in
section \ref{s5}.

Let us then generalize our vector fields $N_R \ell^a$ as follows:
use, in place of $R$, a general function $r (R)$. Recall first
that $N_R$ satisfies $D_a R = N_R \rh_a$ so that we have $N_R d^3V
= dR d^2V$. Therefore,  for more general functions $r(R)$ which
are constant on each leaf $S$ of the foliation, we are led to
choose $N_r$ through $D_a r = N_r \rh_a$. If we use a different
radial function $r^\prime$, then the lapse is rescaled according
to the relation
\begin{equation} N_{r^\prime} =  \frac{dr^\prime}{dr} \, N_r\, .
\end{equation}
Thus, although the lapse itself will in general be a function of
all three coordinates on $H$, the \emph{relative factor} between
any two permissible lapses can be a function only of $r$. This is
the simplest generalization that seems appropriate to the
transition from isolated to dynamical horizons.

Given a lapse $N_r$, following the terminology used in the
isolated horizon framework, the resulting vector fields by
$\xi^a_{(r)}:= N_r\l^a$ will be said to be \emph{permissible}.
Thus, $\xi^a_{(R)}$ used in section \ref{s3.1} is just one
permissible vector field which (on dimensional grounds) happens to
be the convenient one to relate the change $R_2-R_1$ in the
horizon radius to the flux of energy across $\Delta H$. By
repeating the calculation of section \ref{s3.1}, it is easy to
arrive at a generalization of (\ref{ab2}) for \emph{any}
permissible vector field:
\be \label{ab5} \left(\frac{r_2}{2G}- \frac{r_1}{2G}\right) =
\int_{\Delta H} \T_{ab}\th^{\,a}\xi_{(r)}^b\,d^3V \, +\,
\frac{1}{16\pi G} \int_{\Delta H} N_r\left\{ |\sigma|^2 +
2|\zeta|^2\right\} \,d^3V \, , \ee
where the constants $r_1$ and $r_2$ are values the function $r$
assumes on the fixed cross-sections $S_1$ and $S_2$. (Note,
incidentally, that the lapse $N_r$ may well vanish on open
regions. It may also be negative in which case we would have $r_2<
r_1$.) This generalization of (\ref{ab1}) will be useful in
section \ref{s6}.

Here, we simply note a special case of physical interest: $r =
4\pi R^2$. In this case, (\ref{ab5}) directly gives us a formula
for the change in the horizon area (rather than in the horizon
radius):
\be \label{ab6} \left(\frac{a_2}{4G}\, -\, \frac{a_1}{4G} \right)
= \frac{1}{2}\, \int_{\Delta H} \T_{ab}\th^{\,a}\xi_{(r)}^b\,d^3V
\, +\, \frac{1}{32\pi G } \int_{\Delta H} N_r\left\{ |\sigma|^2 +
2|\zeta|^2\right\} \,d^3V \, , \ee
Note however that, as is expected from dimensional reasons, the
right hand side does \emph{not} have the interpretation of the
energy flux across $\Delta H$ even in the case $\Lambda = 0$.
However, since black hole thermodynamics tells us that the
(leading contribution to the) entropy is given by $a/4\ell_{\rm
Pl}^2$, one may wish to interpret the right hand side as the
\emph{entropy flux} through $\Delta H$ (in the $\hbar =1$ units).

\emph{Remark:} In the definition of a dynamical horizon, we
required $\Theta_{(n)} <0$ which guaranteed that $|D R| \not=0$,
i.e., that $R$ is a good coordinate on $H$. This was used in the
derivation of (\ref{ab1}) in section \ref{s3.1}. However, we can
weaken the definition and ask only that $\Theta_{(n)} \le 0$. In
this case, we can introduce a function $x$ such that the
marginally trapped 2-surfaces are labelled by $x = {\rm const}$
and $|Dx| \not =0$ and repeat the calculations of section
\ref{s3.1} to obtain the analog of (\ref{ab1}) in which $R$ is
replaced by $x$. We can then note that although $R$ need not be a
good coordinate on $H$, it is nonetheless a smooth function of the
coordinate  $x$ whence, the calculation of this sub-section can be
repeated to obtain the area balance (\ref{ab1}). Thus, the area
balance law holds also under the weaker assumption $\Theta_{(n)}
\le 0$. If $\Theta_{(n)} >0$, we can reverse the argument to get
an area decrease law appropriate for white holes.

\section{Angular momentum}
\label{s4}

To obtain the integral version of the first law (\ref{1law1}), we
need the notion of angular momentum and angular momentum flux. It
turns out that the angular momentum analysis is rather straight
forward and is, in fact, applicable to an arbitrary space-like
hypersurface. Fix \emph{any} vector field $\fie^a$ on $H$ which is
tangential to all the cross-sections $S$ of $H$. Contract  both
sides of (\ref{momconstr}) with $\fie^a$. Integrate the resulting
equation over the region $\Delta H$, perform an integration by
parts and use the identity $\Lie_\fie q_{ab} = 2D_{(a}\fie_{b)}$
to obtain
\be\label{balanceJ} \frac{1}{8\pi G}\oint_{S_2}K_{ab}
\fie^a\rh^{\,b} \, d^2V - \frac{1}{8\pi G}
\oint_{S_1}K_{ab}\fie^a\rh^{\,b} \, d^2V  
= \int_{\Delta H} \left( T_{ab}\th^{\,a}\fie^b + \frac{1}{16\pi G}
P^{ab}\Lie_\fie q_{ab}\right)\, d^3V\ee
where, as before, $P^{ab}:=K^{ab}-Kq^{ab}$. (Note that we could
replace $\T_{ab}$ with $T_{ab}$ because $g_{ab}\th^{\,a}\fie^b
=0$. Thus the cosmological constant plays no role in this
section.) It is natural to identify the surface integrals with the
generalized angular momentum $J^{\fie}$ associated with
cross-sections $S$ and set
\begin{equation} \label{jdynamic1}J_S^{\fie} =
-\frac{1}{8\pi G} \oint_{S} K_{ab} \fie^a\rh^{\,b} \, d^2V\, ,
\end{equation}
where we have chosen the overall sign to ensure compatibility with
conventions normally used in the asymptotically flat context. The
term `generalized' emphasizes the fact that the vector field
$\fie^a$ need not be an axial Killing field even on $S$; it only
has to be tangential to our cross-sections.

The flux of this angular momentum due to matter fields and
gravitational waves are respectively
\begin{eqnarray} \mathcal{J}^{\fie}_{\m} &=& -\int_{\Delta H}
 T_{ab}\th^{\,a}\fie^b\, d^3V \, ,\\
 \mathcal{J}^{\fie}_{\g} &=& -\frac{1}{16\pi G}
 \int_{\Delta H} P^{ab}\Lie_\fie q_{ab}\, d^3V \, , \end{eqnarray}
and we get the balance equation
\begin{equation}  J_{S_2}^{\fie} - J_{S_1}^{\fie} =
 \mathcal{J}^{\fie}_{\m}+
\mathcal{J}^{\fie}_{\g}\, . \end{equation}
As expected, if $\fie^a$ is a Killing vector of the three-metric
$q_{ab}$, then the gravitational angular momentum flux vanishes:
$\mathcal{J}^{\fie}_{\textrm{g}} = 0$.  For the discussion of the
integral version of the first law, it is convenient to introduce
the \emph{angular momentum current}
\be{j}^{\fie}:=-K_{ab}\fie^a\rh^{\,b}\ee
so that the angular momentum formula becomes
\be J_S^{\fie}= \frac{1}{8\pi G}\, \oint_S {j}^{\fie}\, d^2V\,
.\ee
We conclude with four remarks:
\begin{description}
\item[i.] \emph{Interpretation of $\zeta^a$:} We can use the
expression of $J_S^{\fie}$ to interpret the vector field $\zeta^a$
which features in the gravitational energy flux: $\zeta^a =0$ on
$H$ if and only if $J_S^{\fie} =0$ for \emph{every} $\fie^a$ which
is divergence-free (i.e. preserves the volume-element) on $S$.

\item[ii.] \emph{Relation to other expressions:} Let us restrict
ourselves to vector fields $\fie^a$ which are divergence-free on
each cross-section $S$. The angular momentum $J^{\fie}_S$
associated with these $\fie^a$ have the following interesting
property. Fix a cross-section $S$ of $H$ and consider an
asymptotically flat, partial Cauchy surface $M$ in the space-time
$\man$ with inner boundary $S$. Denote its Cauchy data by
$(\bar{q}_{ab}, \bar{K}_{ab})$. Then, we can extend $\fie^a$ to a
vector field $\phi^a$ which is an asymptotic rotational symmetry
of $(M, \bar{q}_{ab})$ and repeat the above calculation by
replacing $\Delta H$ with $M$.  The surface integral at infinity
is then the standard ADM angular momentum associated with
$\phi^a$. The angular momentum assigned to $S$ is:
\begin{equation} \label{jdynamic2}\bar{J}_S^{\fie} =
-\frac{1}{8\pi G} \oint_{S} \bar{K}_{ab} \fie^a\bar{r}^{\,b} \,
d^2V\, ,
\end{equation}
where $\bar{r}^a$ is the unit normal to $S$ in $M$. By expressing
$K_{ab}$ and $\bar{K}_{ab}$ in terms of $\nabla_a \ell_b$ and
$\nabla_a n_b$, it is straightforward to show that $J_S^{\fie} =
\bar{J}_S^{\fie}$. Thus, given a divergence-free $\fie^a$ on $S$,
the notion of angular momentum associated with $S$ is unambiguous.
Finally, if $\varphi^a$ is the restriction to $S$ of a space-time
Killing vector defined in a neighborhood of $S$, one can define
the angular momentum via Komar integral and it agrees with
$J^\varphi_S$.
\item[iii.] \emph{Dependence on $\varphi^a$:}  In the above
calculation we did not assume that $\fie^a$ is a Killing field on
$H$. However, $J_S^{\fie}$ would represent the \emph{physical}
angular momentum at the `instant' $S$ only if $\fie^a$ is a
Killing field of at least $(S, \q_{ab})$. Suppose $\fie^a$ has
this property both on $S_1$ and $S_2$, but not on all of $\Delta
H$. Still, because of the balance law (\ref{balanceJ}), the total
flux is well-defined and is in fact independent of the way in
which $\fie^a$ is extended off $S_1$ and $S_2$.
\item[iv.] \emph{Gauge fields:} We indicated in section \ref{s2.3}
that there are subtleties associated with gauge fields.
Considerations of angular momentum illustrate this point. In the
above treatment, we just interpreted $\int_{\Delta H}
T_{ab}\th^{\,a}\fie^b\, d^3V $ as the flux of matter angular
momentum across $\Delta H$. But a priori there is some freedom to
shuffle terms between the 3-dimensional flux integrals and the
2-dimensional `angular momentum charge' integrals. Our choice
ensures that, as at infinity, the 2-sphere `angular momentum
charge' integrals $J_S^{\fie}$ depend only on geometric fields and
not on matter. However Hamiltonian considerations often show that,
in order for angular momentum to be the generator of rotations on
the phase space, such a reshuffling is in fact necessary in the
case of gauge fields . Thus, the `angular momentum charge'
integral can in fact depend on gauge fields as well. (In the case
of isolated horizons, this is demonstrated in detail in
\cite{abl2}.) The required shuffling will not affect any of the
equations but would change interpretations of terms in presence of
gauge matter fields.
\end{description}

\section{Integral version of the first law and the horizon mass}
\label{s5}

This section is divided in to three parts. In the first we obtain
an integral generalization of the first law (\ref{1law1}). In the
second, we restrict ourselves to axi-symmetric dynamical horizons
and introduce, for each cross-section $S$, a canonical notion of
energy (which may be interpreted as the instantaneous mass) and
derive a balance law. In the third, we discuss the
distinction between laws of `black hole mechanics' and of `black
hole thermodynamics'.

\subsection{Generalization of the first law of black hole mechanics}
\label{s5.1}

Let us now combine the results of sections \ref{s3} and \ref{s4}
to obtain the \emph{physical process version} of the first law on
$H$. As in section \ref{s3.2}, we will first consider the case
$\Lambda =0$ and then comment on the role played by the non-zero
cosmological constant.

To begin with, let us ignore angular momentum and consider the
vector field $\xi^a_{(R)}$ of section \ref{s3.1}. For each
cross-section $S$ of $H$, there is a well-defined notion of
horizon energy $E^{\xi_{(R)}}(S)$ (given just by the Hawking
mass). Because of the influx of matter and gravitational energy,
$E^{\xi_{(R)}}$ will change by an amount $\Delta\, E^{\xi_{(R)}} =
\mathcal{F}^{(R)}_{\m}+ \mathcal{F}^{(R)}_{\g}$ as we move from a
cross-section $S_1$ to another cross-section $S_2$. Then, the
infinitesimal form of (\ref{ab3}),
\be \frac{dR}{2G} = dE^{\xi_{(R)}}\, ,\ee
suggests that we define \emph{effective surface gravity} $\k_R$
associated with $\xi^a_{(R)}$ as
\be \k_R:= \frac{1}{2R} \ee
so that the infinitesimal expression is recast into the familiar
form
\be \label{1law2} (\frac{\k_R}{8\pi G})\, da = d E^{\xi_{(R)}}\ee
where $a$ is the area of a generic cross-section. (This conclusion
could also have been reached from (\ref{ab6})). For a general
choice of the radial function $r$, the infinitesimal version of
(\ref{ab5}) yields a generalized first law%
\footnote{$E^{\xi_{(r)}}$ has the dimension of energy only if $r$
has the same dimension as $R$. In the following discussion, we
will assume this to be the case.}:
\be \label{1law3} \frac{\k_r}{8\pi G} \,\,da = d E^{\xi_{(r)}} \ee
provided we define the effective surface gravity $\k_r$ of
$\xi^a_{(r)}$ by
\begin{equation} \k_r = \frac{dr}{dR}\,\, \k_R \qquad \textrm{where}
 \qquad \xi^a_{(r)} = N_r\, \ell^a = \frac{dr}{dR}\,\,\xi^a_{(R)} \,
 .\end{equation}
Note that this rescaling freedom in surface gravity is completely
analogous to the rescaling freedom which exists for Killing
horizons, or, more generally, isolated horizons \cite{afk,abl2}.
There, on the horizon $\ell^a$ can be rescaled by a constant and
surface gravity rescales by the same constant. The new feature in
the present case is that we have the freedom to rescale  $N_R\l^a$
and the surface gravity  by a \emph{function of the radius} $R$
rather than just a constant. This is just what one would expect in
a dynamical situation since $R$ plays the role of time along
$H$. Finally, note that the differentials appearing in
(\ref{1law3}) are the \emph{actual variations} of physical
quantities along the dynamical horizon due to an infinitesimal
change in $r$. This is to be contrasted with derivations of the
first law based on phase space variations \cite{rw,afk,abl2},
where one compares quantities defined on distinct (isolated or
Killing) horizons belonging to distinct space-times. Since
quantities defined in distinct equilibrium configurations are
compared, there one obtains a passive form of the first law. By
contrast, (\ref{1law3}) is an active or a \emph{physical process  version}
of the first law. Hence (\ref{ab5}) is a \emph{finite version} of
the first law in absence of rotation. As in the case of isolated
horizons \cite{afk}, even in absence of rotation, there are many
permissible vector fields and each gives rise to a first law.

Next, let us include rotation. As discussed in section \ref{s3.3},
the general strategy is motivated by the isolated horizon
framework. Pick a vector field $\fie^a$ on $H$ such that $\fie^a$
is tangent to the cross-sections of $H$, has closed orbits and has
affine length $2\pi$.%
\footnote{More precisely, $\fie^a$ is a globally defined Killing
field for \emph{some} metric ---not necessarily the physical one,
$\q_{ab}$--- on each 2-sphere cross-section $S$ of $H$.}
(At this point, $\fie^a$ need not be a Killing vector of
$q_{ab}$.) The isolated horizons considerations suggest that it is
now appropriate to replace $\xi^a_{(r)}$ by vector fields $t^a$
which are of the form $t^a=N_r\l^a-\Omega\fie^a$ where $N_r$ is a
permissible lapse associated with a radial function $r$ and
$\Omega$ an arbitrary function of $R$. (On an isolated horizon,
the analogs of these two fields are constants.) Such vector fields
$t^a$ will be said to be \emph{permissible.} Let us now evaluate
the quantity $\int_{\Delta H} T_{ab}\th^{\,a}t^b\, d^3V$ by taking
a linear combination of (\ref{ab5}) and (\ref{balanceJ}). We
obtain:
\begin{eqnarray}  &&\frac{r_2-r_1}{2G} + \frac{1}{8\pi G} \biggl\{
\oint_{S_2}\Omega j^\fie\,d^2V -\oint_{S_1} \Omega j^\fie\,d^2V
\biggr. - \biggl.\int_{\Omega_1}^{\Omega_2} d\Omega \oint_S
j^\fie\,d^2V \biggr\} = \nonumber\\
&& \int_{\Delta H} T_{ab}\th^{\,a}t^b \,d^3V + \frac{1}{16\pi
G}\int_{\Delta H} N_r \left(|\sigma|^2 + 2|\zeta|^2\right)\,d^3V -
\frac{1}{16\pi G} \int_{\Delta H}\Omega P^{ab}\Lie_\fie q_{ab} \,
d^3V \label{balance1}
\end{eqnarray}
\emph{These are our balance equations in presence of angular
momentum.} There are infinitely many balance equations because
there are infinitely many permissible vector fields. In section
\ref{s5.2}, we will show that, when the horizon metric $q_{ab}$ is
axi-symmetric, one can choose a preferred vector field $t_o^a$
(which is adapted to the Kerr time-translation Killing field in a
precise sense.) For this vector field, given a cross-section $S$,
we will provide an explicit expression of the energy $E^{t_o}_S$
such that the left side of Eq. (\ref{balance1}) can be
re-expressed as the difference $E^{t_o}_{S_2} - E^{t_o}_{S_1}$,
whence we are led to a preferred balance equation:
\ba \label{balance2} E^{t_o}_{S_2} - E^{t_o}_{S_1} =  \int_{\Delta
H} T_{ab}\th^{\,a}t^b_o \,d^3V &+& \frac{1}{16\pi G}\int_{\Delta
H} N_o \left(|\sigma|^2 + 2|\zeta|^2\right)\,d^3V\nonumber\\
& -& \frac{1}{16\pi G} \int_{\Delta H}\Omega_o P^{ab}\Lie_\fie
q_{ab} \, d^3V \,. \ea

Let us return to the general case considered in  Eq.
(\ref{balance1}). Assuming there is a well-defined notion $E^t$ of
the horizon energy at each cross-section, with the right side of
(\ref{balance1}) its flux, we can now obtain the first law for
mechanics for dynamical horizons. Let us restrict ourselves to
infinitesimal $\Delta H$. Then, the three terms in the curly
brackets combine to give $d(\Omega J) - J d\Omega$ and eq.
(\ref{balance1}) reduces to
\begin{equation} \label{1law4}  \frac{dr}{2G}+
\Omega dJ \equiv \frac{\k_r}{8\pi G}da + \Omega dJ\,= \, dE^t .
\end{equation}
This is just the familiar first law but now in the setting of
dynamical horizons. Since the differentials in this equation are
variations of physical quantities along $H$, this can be viewed as
a \emph{physical process version} of the first law of black hole
mechanics. Note that for each allowed choice of lapse $N_r$,
angular velocity $\Omega(r)$ and vector field $\fie^a$ on $H$, we
obtain a permissible time vector field
$t^a=N_r\l^a-\Omega\fie^a$ and \emph{a corresponding first law}.
For isolated horizons \cite{afk,abl2} the situation is similar;
there are infinitely many permissible vector fields and a first
law for each of them. The main difference is that we are now in a
dynamical situation and  (\ref{1law4}) tells us what happens
instantaneously on the dynamical horizon (at the `instant'
represented by the cross-section $S$). The first law in
\cite{afk,abl2} describes transitions from one equilibrium
situation to a nearby one and refers to the isolated horizon as a
whole. Again, the generalization from that time independent
situation consists of allowing the lapse and the angular velocity
to become $R$-dependent, i.e., dynamical. Therefore, for vector
fields $t^a$ for which there is a satisfactory notion of horizon
energy $E^t$ (as for $t^a=t^a_o$ introduced in section
\ref{s5.2}), eq. (\ref{balance1}) yields an \emph{integral,
physical process version generalization of the familiar,
differential first laws of isolated horizon mechanics}:
\be \label{integral1law} \frac{r_2-r_1}{2G} + \frac{1}{8\pi G}
\biggl\{ \oint_{S_2}\Omega j^\fie\,d^2V -\oint_{S_1} \Omega
j^\fie\,d^2V \biggr. - \biggl.\int_{\Omega_1}^{\Omega_2} d\Omega
\oint_S j^\fie\,d^2V \biggr\} = E^{t}_{S_2} - E^{t}_{S_1} \ee

Thus, if a suitable notion of horizon energy $E^t_S$ can be found,
the same equation (\ref{balance1}) can be used to obtain an energy
balance equation (\ref{balance2}) similar to that of Bondi and
Sachs, but now at the dynamical horizon, \emph{and} an integral
generalization (\ref{integral1law}) of the active form of the
first law of black hole mechanics.

Finally, let us consider the case when the cosmological constant
is non-zero. Then, the integral version of the first law is given
simply by replacing $T_{ab}$ in eq. (\ref{balance1}) by
$\bar{T}_{ab}$. In the infinitesimal version, we now obtain
\begin{equation} \label{1law5}  \frac{\k_r}{8\pi G}[ 1- \Lambda R^2]\,
da + \Omega dJ\,= dE^t\, .
\end{equation}
Thus, the only effect that a cosmological constant has is to
modify the expression of effective surface gravity. This is
completely analogous to what happens to the standard first law on
Killing or isolated horizons.

We conclude with two remarks:
\begin{description}
\item[i.] \emph{More general permissible vector fields:} Since we
selected the vector fields $t^a$ using intuition
derived from isolated horizons, we were led to ask that $N_r/N_R$
and $\Omega$ be  functions only of $R$. But it is rather easy to
allow more general $N, \Omega$ and thus extend the notion of
permissible vector fields.  Set $\bar{t}^a = N\ell^a - \Omega
\varphi^a$, where $N$ is
\emph{any} smooth function on $H$, not necessarily tied to a
radial function $r$. Then, we obtain an obvious generalization of
the balance equation (\ref{balance1}). Furthermore, we can set the
effective surface gravity to be
\be \k_{\bar{\xi}} = \left(\frac{1}{8\pi}\oint N N_R^{-1}\,
\twoR\,\, d^2V  \right)\, \k_R \ee
and again obtain the first law (\ref{1law4}) with $\kappa_r$
replaced by $\kappa_{\bar{\xi}}$.

Our restriction on $\Omega$ being only a function of $R$ corresponds
to considering rigidly
rotating fields $t^a$ (where, however, the angular speed of
rotation is allowed to vary as one moves from one cross-section to
another). This restriction is necessary to recover the familiar
infinitesimal form of the first law and also for the definition of
the horizon energy in section \ref{s5.2}. However, as far as the
integral first law is concerned, one can easily accommodate
differential rotation by allowing $\Omega$ to be a function also
of angular coordinates.
\item[ii.] \emph{The $\Theta_{(n)}\leq 0$ case:} It is easy to
verify that the main result of this section goes
through even if the condition $\Theta_{(n)} < 0$ is weakened to
allow $\Theta_{(n)} \le 0$. The reasoning is the same as that in
the remark at the end of section \ref{s3}.
\end{description}

\subsection{Horizon mass}
\label{s5.2}

Recall first the situation at null infinity. Given a time
translation ${t}$ in the Bondi-Metzner-Sachs group and a
cross-section $\mathbb{S}$ of $\mathbb{I}^+$, we can define
Bondi-energy $E^{t}_S$ such that the difference between the energy
associated with any two cross-sections equals the Bondi-flux
through the region of $\mathbb{I}^+$ they bound
\cite{bondiflux,as,wz}. On dynamical horizons, the right side of
eq. (\ref{balance1}) provides us with the analog of the
Bondi-flux. It is natural to ask if there is also a
\emph{satisfactory} notion of energy $E^t_S$ associated with each
cross-section $S$. In this sub-section, we will address this issue
using dual considerations: finding preferred fields for which a
mathematically viable notion of $E^t_S$ exists \emph{and} admits a
satisfactory physical interpretation. We will first restrict
ourselves to the case $\Lambda =0$ and show that the both goals
can be met for axi-symmetric dynamical horizons.

Given any permissible vector field $t^a$ on $H$, we can just solve
the ordinary differential equation on $H$,
\be \label{1law6}\frac{dE^t}{dR} =  \, \frac{R}{G}\, \k_r (R)\, +
\Omega\, \frac{dJ}{dR}\, ,\ee
derived from (\ref{1law4}), and obtain an expression $E^t_S$ on
any cross section $S$. But in general the result will not be
expressible in terms of geometric quantities defined locally on
$S$. If it is, we will have a mathematically viable notion of
$E^t_S$. Our second requirement is that the resulting $E^t_S$
should have a direct physical interpretation.

The first example is provided by dynamical horizons $H$ on which
the intrinsic metric $q_{ab}$ of $H$ is spherically symmetric.
Then, it is natural to choose $\Omega=0$ and $R$ as the radial
coordinate so that the preferred vector field is $t_o^a = N_R
\ell^a$ with effective surface gravity $\k_R = 1/2R$. In this
case, the integration of the flux yields:
\be \label{hm} E^t_S = \frac{R}{2G}\, , \ee
where the integration constant has been chosen such that $E^t_S$
tends to the isolated horizon mass when the matter flux vanishes
and the horizon reaches equilibrium. Since we arrived at this
expression by integrating the differential equation (\ref{1law4}),
and since the right side of this equation does not refer to matter
fields at all, the expression of $E^t_S$ is purely geometric. In
fact, since the expansion $\Theta_{(\ell)}$ vanishes on $S$, as
noted before, $E^t_S$ is precisely the Hawking mass of $S$. In the
spherically symmetric case, this is a physically viable measure of
energy in the black hole; thus both our goals are met.
Furthermore, by restricting the balance law (\ref{balance1}) to
this case, we conclude:
\be E^t_{S_2} - E^t_{S_1} = \int_{\Delta H} T_{ab}\th^{\,a}t^b
\,d^3V. \ee
Thus, (\ref{balance1}) has a clear-cut interpretation in this
case: the flux of gravitational energy vanishes, and the increase
in $E^t(S)$ is fully accounted for by the matter flux
$\mathcal{F}^{(R)}_{\m}$. Note that this was obtained assuming
spherical symmetry \emph{only of} $(H, q_{ab})$.

Beyond spherical symmetry, the gravitational energy flux would not
be zero, whence the balance equation will be non-trivially
generalized. We can begin with distorted but non-rotating
dynamical horizons, i.e., ones on which the angular momentum
current density $\q^{ab}K_{bc}\rh^c$ vanishes. Again, it is
appropriate to set $\Omega=0$.  Furthermore, from isolated horizon
considerations, we know that the distortion does \emph{not} affect
surface gravity \cite{afk}. Therefore we can again set $r =R$.
Thus, the discussion is reduced to that in the spherically
symmetric case. Again, the isolated horizon framework supports the
interpretation of the Hawking mass as the horizon mass in this
case as well. The difference from spherical symmetry is that now
there may be gravitational radiation. Thus, in the distorted case,
the balance equation derived from (\ref{balance1}) is more
general:
\be E^t_{S_2} - E^t_{S_1} = \int_{\Delta H} T_{ab}\th^{\,a}t^b
\,d^3V + \frac{1}{16\pi G}\int_{\Delta H} N_R \left(|\sigma|^2 +
2|\zeta|^2\right)\,d^3V \, .\ee
(However, because the angular momentum current vanishes, the
expression of $\zeta^a$ simplifies to: $\zeta^a = \q^{ab}D_b\ln\,
N_R$.)

Finally let us incorporate rotation. Physically, the most
interesting case is the one in which $q_{ab}$ is axi-symmetric,
with $\fie^a$ as its axial Killing vector. (In what follows, we
will work with this fixed $\fie^a$. The dependence on $\fie^a$ of
various physical quantities such as the angular momentum will now
be dropped.) To specify a preferred vector field $t_o^a$, we need
to specify $\k_r$ and $\Omega$. The idea is to apply, on each
cross-section $S$ of $H$, the strategy used in the isolated
horizon framework to
select a \emph{preferred} permissible vector field $t_o^a$:\\
i) Calculate the angular momentum $J_S$ defined by the axial
Killing field $\fie^a$. This provides us with a function $J(R)$ on
the horizon $H$; \\
ii) Set
\begin{equation}\label{kerrkappa} \k_r = \kappa_o(R) :=
\frac{R^4-4G^2J^2}{2R^3\sqrt{R^4+4G^2J^2}}. \end{equation}
This is achieved by solving for $dr/dR = 2 R\k_o(R)$, which
determines $r$ and $N_r$; and,\\
iii) choose $\Omega$ such that
\begin{equation}\label{kerromega} \Omega= \Omega_o(R) :=
\frac{2GJ}{R\sqrt{R^4+4G^2J^2}} \, .\end{equation}
This functional dependence of $\k_r$ and $\Omega$ on $R$ and $J$
is exactly that of the Kerr family. That is, given a cross-section
$S$, we choose $t_o^a$ which has the same effective surface
gravity and angular velocity at that cross-section as the surface
gravity and angular velocity that \emph{the} time-translation
Killing field has on the horizon of the Kerr solution with the
same area and angular momentum. Our task now is to integrate eq.
(\ref{1law6}). For this, let us first recall the properties of the
standard Smarr formula for the Kerr family:
\begin{equation}\label{mass} M(R,J)
:= 2\, (\frac{\kappa_o a}{8\pi G} + \Omega_o J) =
\frac{\sqrt{R^4 + 4G^2J^2}}{2GR} \, .\end{equation}
The function $M$ of two variables $R,J$ has the property that
under arbitrary variations of the two parameters the first law,
$\delta M = (\kappa_o/8\pi G) \delta a + \Omega \delta J$, is
satisfied. Therefore, it follows that
\be E^{t_o}(R) := M(R, J(R)) \ee
satisfies the differential equation (\ref{1law6}). Furthermore, it
is the unique solution which reduces to the expression (\ref{hm})
in the case of spherical symmetry (when $J(R) =0$ identically).

This notion of horizon energy has some attractive properties.
First, it depends only on geometrical fields on each cross section
and the dependence is local. Yet, as noted in section \ref{s5.1},
thanks to the constraint part of Einstein's equations, changes in
$E^{t_o}$ over \emph{finite} regions $\Delta H$ of $H$ can be
related to the expected fluxes:
\be E^{t_o}_{S_2} - E^{t_o}_{S_1} = \mathcal{F}^{(t_o)}_\g +
\mathcal{F}^{(t_o)}_\m \, ,\ee
where the flux of gravitational energy $\mathcal{F}^{(t_o)}_\g$ is
local and positive definite (see (\ref{balance2})). (The
gravitational angular momentum flux which, in general, has
indeterminate sign vanishes due to axi-symmetry.) Finally, as
mentioned in section \ref{s3}, Booth and Fairhurst have recently
shown that this expression of the dynamical horizon energy emerges
from a systematic Hamiltonian framework on space-times $\man$ with
a dynamical horizon $H$ as inner boundary \cite{bf2}.

Note that, as a function of its angular momentum and area, each
cross-section $S$ is assigned simply that $E^{t_o}_S$ which it
would have in the Kerr family. Physically, this is a simple and
attractive property. Furthermore, because of its close relation to
the Kerr time translation, $t_o\mid_S$ represents that
`time-translation for which the horizon is at rest at the instant $S$'.
Therefore, we will refer to $E^{(t_o)}$ as the \emph{mass
function} on the (axi-symmetric) dynamical horizon $H$ and set
$E^{t_o} =M(R)$. (The overall strategy is the same as that used in
the isolated horizon framework \cite{abl2}.)  Thus, among the
infinitely many first laws (\ref{1law4}), there is a canonical
one:
\begin{equation}  dM = \frac{\k_o}{8\pi G} da + \Omega_o dJ \, .
\end{equation}

We conclude this section with a discussion of the possibility that a
dynamical horizon can have an excess of angular momentum and violate
the Kerr bound $J\le GM^2$ and the possibility of extracting rotational
energy from the black hole.%
\footnote{ For the discussion that follows, it is convenient to
note that in the Kerr family the limiting, extremal Kerr horizon
results when $2GJ = R^2 = 2G^2M^2$.} %
In the Kerr solution, it is forbidden to violate the inequality
$J\le GM^2$.  However, none of the equations we derived rule out
the possibility that a dynamical horizon may be formed with a
cross section $S$ on which the Kerr limit is violated, i.e., $ 2JG
> R^2$. On this $S$, we will have $\kappa_o <0$ (see eq.
(\ref{kerrkappa})) so that, with our prescription for constructing
$t_o$, $\k_r$ would be negative (whence $r$ would be a decreasing
function of $R$). But the prescription for selecting $t^a_o$ still
goes through and the dynamical horizon mass, given by:
\be M(R) = \frac{\sqrt{R_\Delta^4 + 4 G^2 J^2}}{2G R_\Delta}\ee
which is well-defined, positive. Let us first consider the case
when $T_{ab}$ vanishes on the horizon. Then, because of
axi-symmetry, $J$ is constant. On the other hand, the area always
increases. What happens to surface gravity? In the Kerr allowed
region it is positive and in the Kerr forbidden region,
negative. Are we driven toward the Kerr allowed region or
further away from it? A simple calculation yields:
\be \left(\frac{\partial \kappa_o}{\partial R}\right)_{J}
       \begin{array}{ccl}
         >0 & \textrm{if} & R^2 < R_o^2 \\
         <0 & \textrm{if} & R^2 > R_o^2\, .
       \end{array}
\ee
where $R_o^2 \sim 5.085JG$.
\begin{figure}
  \begin{center}
  \includegraphics[height=7cm]{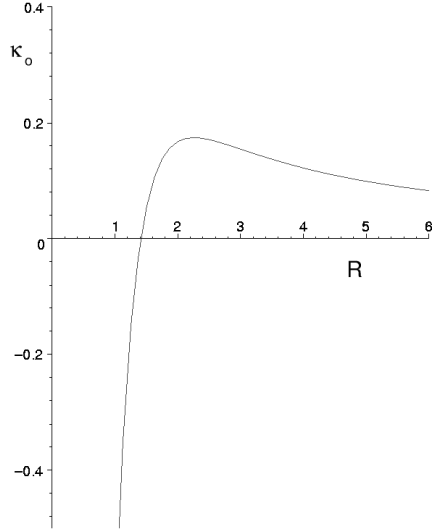}
  \caption{A plot of the Kerr surface gravity $\kappa_o$ (from
   eq. (\ref{kerrkappa})) as a function of $R$ (with $JG$ set equal
   to $1$ for definiteness).  The part $\kappa_o<0$ is the
   \emph{Kerr forbidden} side while $\kappa_o>0$ is the \emph{Kerr
   allowed} regime. Since $R$ increases monotonically with time,
   this graph shows that the dynamical horizon always
   evolves toward the Kerr allowed region under time
   evolution.}\label{figkappa}
  \end{center}
\end{figure}
This is also shown graphically in figure \ref{figkappa} as a plot of
$\kappa_o(R)$ versus $R$ for a fixed value of $J$.
Therefore, on the Kerr-forbidden side, under time evolution
$R$ increases whence the surface gravity also increases, i.e.,
becomes less negative, and we are pushed toward the extremal
point. When the radius increases so that $R^2> 5.085 JG$, surface
gravity starts decreasing but this is essentially irrelevant
because we are now on the Kerr-allowed side where surface gravity is always
positive (and tends to zero as area tends to infinity, keeping $J$
fixed). These considerations suggest that a black hole may well be
formed in the Kerr-forbidden region and then settle down to a
Kerr hole as time evolves.

Since we put $T_{ab} =0$, this process can happen even in vacuum
general relativity, e.g., in black hole mergers. At first this
seems counter-intuitive because there are heuristic arguments
which suggest that the black hole cannot radiate more angular
momentum than energy, whence if it is initially formed with $J >
GM^2$, it would not be able to settle down to a Kerr state in the
distant future. However, what can happen is the following.
Initially, one may have $J> GM^2$ but there may be energy trapped
between the black hole and the `peak of the potential' outside the
horizon, which may fall in the black hole, increasing its mass
significantly but keeping its angular momentum the same, thereby
moving its state toward the Kerr allowed regime. Finally, in
presence of matter, the flux $\mathcal{F}^{(t_o)}_\m$ need not be
positive definite because if $\Omega\not= 0$, the vector field
$t^a_o$ is space-like at the horizon. In this case, the in-falling
matter could pour negative angular momentum into the black hole
thereby decreasing its mass.

Thus, it is rather surprising that there is no obvious obstruction
for a dynamical horizon to be first formed in the Kerr forbidden
region and yet fulfill the physical expectation that the final,
equilibrium state should be a Kerr horizon. We should emphasize
however that the issue of whether this is compatible with
solutions to the constraint equations on $H$ is yet to be
analyzed. In physically interesting situations, there is a further
very non-trivial restriction: one is interested only in those
horizons which arise in the dynamical evolution of physically
appropriate initial data on Cauchy surfaces. Nonetheless, the fact
that there is no obvious obstruction suggests that the issue
should be analyzed further.

We conclude with two remarks.
\begin{description}
\item[i.] \emph{The cosmological constant:} Our discussion can be
generalized to the $\Lambda\not=0$ case
in a rather straightforward manner by replacing the current, Kerr
expressions of $\kappa_o(R)$ and $\Omega_o(R)$ by those from
Kerr-de Sitter and Kerr-anti de Sitter space-times.
\item[ii.] \emph{Physical relevance of $t_o^a$:} In this sub-section,
we introduced a family of physically
motivated vector fields $t_o^a$ and showed that the corresponding
energy $E^{t_o}_S$ is determined by fields defined locally on $S$.
However, this is by no means the only vector field with this
property. To a certain extent, an analogy at null infinity is
provided by the Bondi-Metzner-Sachs supertranslations: There is an
infinite dimensional family of supertranslations, each associated
with a local flux, a 2-sphere supermomentum integral, and a
balance law \cite{as}. Although the supermomenta and their fluxes
do carry physical information, it is the 4-momentum and its flux
that is most important physically and admits a direct and
transparent interpretation. Similarly, on dynamical horizons, of
all permissible vector fields $t^a$ leading to local energy
expressions $E^t_S$, it is likely that $t^a_o$ would be the most
relevant one from physical considerations. In particular, one
expects that in the asymptotic future the dynamical horizon would
tend to a Kerr isolated horizon \cite{abl2,lp} and $E^{t_o}_S$
would tend to the mass of that Kerr space-time.
\end{description}

\subsection{Mechanics versus thermodynamics}
\label{s5.3}

In stationary space-times ---and more generally, in the isolated
horizon framework--- the horizon geometry is time independent and
this in particular implies that the surface gravity $\kappa$ is
constant on the horizon. In the physical process version of the
first law, $dE = (\kappa/8\pi G)da + \Omega dJ$, one considers
transitions from a time independent state to a nearby time
independent state. Conceptually, this is the same setting as in
laws of equilibrium thermodynamics. As in the case of the first
law of thermodynamics, the second term represents mechanical work
done on the horizon while the first term does not; it is
interpreted as the analog of the term $TdS$ representing the `heat
absorbed by the black hole'. The specific form of this term shows
that, in infinitesimal processes involving black holes, the change
in surface gravity can be ignored just as the change in the
temperature is ignored in the transitions envisaged by the first
law.

By contrast, in this paper we considered fully dynamical
situations in which the horizon geometry can be very far from
being stationary. We obtained two closely related results, the
balance equation (\ref{balance2}) and the integral generalization
(\ref{integral1law}) of the first law. The first can be directly
interpreted as a statement of conservation of energy, in that it
describes how the energy of the dynamical horizon changes because
of the influx of matter and gravitational waves. The second is an
integral version of the first law of black hole mechanics because
it tells us how the changes in the characteristics of the black
hole --- the area and the angular momentum ---  are correlated with
changes in its energy.

Let us focus on the second. The angular momentum term can again be
interpreted as mechanical work done on the black hole. What about
the term representing the change in area? Is there again a close
analogy with thermodynamics? To analyze this issue, we must
consider fully \emph{non}-equilibrium thermodynamical processes.
Generically, the system does not have time to come to equilibrium
in these processes and there is no canonical notion of its
temperature. Therefore, while one can still interpret the
difference $E_2 - E_1 - ({\rm work})$ as the heat absorbed by the
system, in general there is no longer a clean split of this term
in to a temperature part and a change in entropy part. If the
process is such that the system remains close to equilibrium
throughout the process, i.e., can be thought of as making
continuous transitions between a series of equilibrium states,
then the difference can be expressed as $\int T dS$, where the
temperature $T$ varies slowly during the transition.

The situation on dynamical horizons is analogous.%
\footnote{This point was emphasized by S. Fairhurst at the  Black
hole IV workshop, held at Honey Harbor, Canada in May 2003 and at
the Penn State Decennial conference in June 2003.} %
On general dynamical horizons, the time dependence can be strong,
and physically one does not expect the term $E^{t_o}_{S_2} -
E^{t_o}_{S_1} - ({\rm work})$ to admit a natural split into a
temperature part and a change in entropy part. Indeed, if
the horizon geometry is changing very rapidly, it cannot be
considered to be in a near-equilibrium state whence it would be
inappropriate to associate an instantaneous physical temperature
to it. How does one reconcile this with the fact that in eq.
(\ref{balance1}) the difference \emph{is} expressed as $\int \k_r
da$ ? The resolution lies in the fact that $\k_r$ is only the
\emph{effective} surface gravity. More precisely, in striking contrast
to what happens in \emph{equilibrium} configurations represented
by isolated horizons, $\k_r$ does not have the \emph{geometrical}
interpretation of surface gravity; as shown in section \ref{s6},
it can only be interpreted as the 2-sphere average of a
geometrical surface gravity associated with certain vector fields
on $H$. This is a reflection of the limitation that, in highly
dynamical situations, $\k_r$ should not have a direct
interpretation of instantaneous, physical temperature.

One would expect such an interpretation to be meaningful only if
the time dependence is weak, i.e., on weakly-dynamical horizons
which can be regarded as perturbations of isolated horizons. In
this case, the geometrical surface gravity \cite{bf2} (see
section \ref{s6}) would be approximately constant and thus
approximately equal its average, $\k_r$. In this situation, one
can regard the horizon as making continuous transitions from one
equilibrium state to another and then the geometrical surface
gravity appears to be a good analog of the (slowly varying)
temperature.  In these situations, the dynamical first law
(\ref{balance1}) can be simplified by keeping terms only up to
second order in perturbations away from isolation \cite{bf1}. In
this approximation, $E^{t_o}_{S_2} - E^{t_o}_{S_1} - ({\rm work})$
can be interpreted as `$\int \kappa da$' where $\kappa$ has a
geometrical interpretation as surface gravity. Hence, the
simplified version of (\ref{integral1law}) can be regarded as the
integral version of the first law of black hole
\emph{thermodynamics}.

These considerations have interesting implications to the notion
of black hole entropy in dynamical situations. Because the horizon
area increases in dynamical processes, in view of the second law
of thermodynamics, it is tempting to identify a suitable multiple
the horizon area with entropy. In time independent situations,
this interpretation is confirmed also by the first law
(\ref{1law1}) because the term $(\kappa/8\pi G)\, da$ is analogous
to the term $TdS$ in the first law of thermodynamics. The above
discussion suggests that the interpretation should continue to be
valid also on weakly-dynamical horizons. It is therefore
interesting to analyze if the black hole entropy derivations based
on counting of micro-states, such as those of \cite{abck}, can be
extended to this case. For highly dynamical situations, on the
other hand, the situation is less clear. In the case of event
horizons, for example, one would not expect this formula for
entropy to be meaningful because, as mentioned in section
\ref{s1}, an event horizon can be formed and grow in a flat space
region in anticipation of a future gravitational collapse. It is
difficult to imagine how a quasi-local counting of micro-states
can account for this phenomenon. The case of highly dynamical
horizons falls in-between. On the one hand, the case for
identification of entropy with a multiple of area is now much
weaker than on weakly-dynamical horizons for reasons discussed
above. On the other hand, somewhat surprisingly, the term $(r_2
-r_1)/2G$ of (\ref{integral1law}) \emph{can be} expressed as
$(1/8\pi G)\,\int \k_r da$ even in the fully dynamical regime.
Furthermore, since the growth of area \emph{is} related to the
energy flux across the horizon, it may well be possible that a
quasi-local counting arguments along the lines of \cite{abck} can
be constructed in this case.

\section{Transition to equilibrium}
\label{s6}

The conventions we used in all the calculations up to this stage
are well-suited to the space-like character of $H$. When $H$
reaches equilibrium, there is no longer a flux of matter or
gravitational energy across it whence (with appropriate
normalization) the shear and the matter flux vanishes. Eq.
(\ref{spacelike}) now implies that the horizon must become null.
Furthermore, since the expansion $\Theta_{(\ell)}$ vanishes, it is
a \emph{non-expanding horizon} in the sense of \cite{afk,abl1}.
The goal of this section is to analyze the transition from a
dynamical horizon to a non-expanding one. In most physical
situations, because of back-scattering, one can expect the
equilibrium to be reached only asymptotically, i.e., in the
infinite future. (For exceptions, see the examples discussed in
Appendix \ref{a1}.) However, as we will see, the case of
asymptotic equilibrium is technically simpler but the subtleties
involved in the matching at a finite time are more instructive.

\subsection{The non-expanding horizon}
\label{s6.1}

Let us then consider a 3-manifold $M$, topologically $S^2\times R$
which is the union of a dynamical horizon $H$ and a non-expanding
horizon $\Delta$ (see figure 1). Thus, $H$ is space-like and
foliated by a family of marginally trapped surfaces $S$, while
$\Delta$ is null. Denote the past boundary of $\Delta$ by $S_0$
which will be assumed to be the (uniform) future limit of the
cross-sections $S$ of $H$. We will assume that: i) the space-time
metric $g_{ab}$ is $C^k$ for some $k\ge 2$; ii) $M$ is a $C^{k+1}$
sub-manifold; and iii) the pull-back $q_{ab}$ to $M$ of $g_{ab}$
admits an axial Killing field $\fie^a$.

Let us first consider $\Delta$. It has the property that the
expansion of \emph{any} of its null normals vanishes. However, to
extract physics, one needs to endow it with an additional
structure. Since $\Delta$ is null, it follows that $\lb^a\nabla_a
\lb^b = \kappa_{\lb} \lb^b$ for any of its null normals $\lb^a$.
The extra structure consists of an equivalence class $[\lb]$ of
null normals whose acceleration, or surface gravity
$\kappa_{\lb}$, is \emph{constant} on $\Delta$, where $\lb^a$ and
$(\lb^\prime)^a$ are equivalent if and only if $(\lb^\prime)^a = c
\lb^a$ where $c$ is a constant on $\Delta$. Such a choice can
\emph{always} be made but it is far from being unique \cite{abl1}.
(The freedom is exhibited in section \ref{s6.2}.) The pair
$(\Delta, [\lb])$, where $[\lb]$ satisfies this condition, defines
a \emph{weakly isolated horizon}.

On weakly isolated horizons, one can introduce the notion of
energy and angular momentum such that the zeroth and the first
laws of black hole mechanics hold. Not only is the angular
momentum $J_\Delta^\fie$ conserved as one would expect because of
`isolation', but its value turns out to be independent of the
specific choice of $[\lb]$ made in the transition from the
non-expanding horizon to the weakly isolated one. To define
energy, one needs to introduce `permissible' vector fields $t^a$
\cite{afk,abl2} and there is an infinite family of these. However,
on any weakly isolated horizon $(\Delta, [\lb])$, one can choose a
canonical one, $t_0^a = \lb_0^a - \Omega_0 \fie^a$, such that the
surface gravity of $\lb_0^a$ and the angular velocity $\Omega_0$
are determined by the area $a_\Delta$ and the angular momentum
$J^\fie_\Delta$ exactly as on the Kerr horizon. Again, when this
choice is made, the value of $E^{t_0}_\Delta$ is independent of
the specific equivalence class $[\lb]$ chosen in the transition
from non-expanding to weakly isolated horizons. In this sense, the
angular momentum and the mass are properties of non-expanding
horizons themselves, although in the intermediate stages in the
calculation one has to pick a weakly isolated horizon structure.

However, the vector fields $t_0^a$ do vary with the choice of
admissible $[\lb]$; they all just happen to lead to the same value
of energy. In our case, $\Delta$ is the limit of a dynamical
horizon $H$, whence it is natural to pick that $t^a_0$ on $\Delta$
which arises as the limit of the canonical vector field $t^a_o$ on
$H$ (introduced in Section \ref{s5.2}). This will in turn fix the
weakly isolated horizon structure on $\Delta$ uniquely.

\subsection{An intermediate construction on the dynamical horizon}
\label{s6.2}

To carry out the matching, we need to introduce some additional
structure on $H$. This structure will enable us to take the limits
to $S_0$ and also clarify the meaning of the `effective surface
gravity' introduced in section \ref{s5.1}.

Throughout our calculations so far, we used the unit normal
$\th^a$ to $H$ and the unit normal $\rh^a$ within $H$ to the
cross-sections $S$. In particular the null vectors $\ell^a, n^a$,
which played a dominant role throughout, are the sum and
differences of these normals. This structure is well-suited to the
space-like character of $H$. However, when we consider the
transition, because $\Delta$ is null, these fields either diverge
or vanish as we approach $S_0$. Therefore, to study the limit to
equilibrium, we need to rescale these fields suitably.

With this goal in mind, let us begin by introducing a smooth
vector field $\Vb^a$ which can be regarded as a smooth
(space-like) extension to $H$ of a suitable $\lb^a$ on $\Delta$.
As discussed in section \ref{s6.1}, we would like $\lb^a$ on
$\Delta$ such that its surface gravity $\kappa_{\lb}$ equals the
Kerr value $\kappa_o(a_\Delta, J_\Delta)$. Such vector fields
always exist and the freedom in their choice is given by
\cite{abl1}:
\be \label{rescaling1} \lb^{\prime a} = (1+ A\, e^{-\kappa_o v})\,
\lb^a \ee
if $\kappa_o\not= 0$ and
\be \label{rescaling2} \lb^{\prime a} = B \lb^a\ee
$\kappa_o =0$, where $v$ is the affine parameter of $\lb^a$ and
the functions $A, B$ on $\Delta$ satisfy: ${\cal L}_{\lb} A =0,\,
{\cal L}_{\lb} B =0$ and $(1 + A) >0, \, B
>0$. On $H$, it is natural to require that $\Vb^a$ to be parallel
to $\rh^a$ and map cross-sections of $H$ to themselves. This
condition determines $\Vb^a$ up to a rescaling by a function of
$R$. Given any one of these $\Vb^a$, one can use the freedom
available in the choice of $\lb^a$ on $\Delta$ to pick one that
will match smoothly with $V^a$ on $H$.

Fix one of these vector fields $\Vb^a$ and define multiples
$\nb^a$ and $\lb^a$ of $\ell^a$ and $n^a$ via: $V^a\nb_a =-2$ and
$\lb^a\nb_a = -2$. By construction, the barred fields are smooth
on $M$. Using the fact that $\Vb^a$ is parallel to $\rh^a$, we
have: $\Vb^a = \lb^a -b^2 \nb^a$ for some smooth function $b$.
Since $\Vb^a \Vb_a = 4b^2$, and since $\Vb^a$ becomes null at
$S_0$, it follows that $b$ tends to zero as we approach $S_0$
along $H$  and stays zero on $\Delta$. It is easy to check that
the null fields $\lb^a$ and $\nb^a$ are related to our original
$\ell^a$ and $n^a$ through $\lb^a = b \ell^a$ and $\nb^a = b^{-1}
n^a$. Since barred fields are smooth across $S_0$, it follows that
$\ell^a$ diverges and $n^a$ tends to zero as we approach $S_0$.
Since $R$ is constant on $\Delta$, we also know that $dR$ and
hence $N_R$ and $N_r$ all go to zero as we approach $S_0$ along
$H$. In section \ref{s6.3} we will show that they do so at the
`same rate' as $b$. Since $\lb^a = b \ell^a$, this will establish
that the field $N_r\ell^a$ used in the construction of $t_o^a$ on
$H$ admits a limit to $S_0$. (Throughout this discussion, $N_r$
will be the lapse featuring in the expression of the canonical
vector field $t_o^a = N_r \ell^a - \Omega(R) \fie^a$ on $H$.)

\subsection{Matching of physical quantities}
\label{s6.3}

Let us begin by showing that the limit of the angular momentum
$J^\fie_S$ of cross-sections $S$ of $H$ equals the angular
momentum $J^\fie_\Delta$ defined on $\Delta$. The angular momentum
on any cross-section $S$ of $H$ is given by (\ref{jdynamic1}):
\begin{equation} {J}_S^{\fie} =
-\frac{1}{8\pi G} \oint_{S} {K}_{ab} \fie^a\rh^{\,b} \, d^2V\, ,
\end{equation}
Using the definition of $K_{ab}$ and expanding the vectors $\th^a$
and $\rh^a$ in terms of $\lb^a$ and $\nb^a$ which are well-defined
on all of $M$, we can  rewrite the integrand as
\be {K}_{ab} \fie^a\rh^{\,b} =  \bar{\omega}_a \fie^a -
\fie^a \nabla_a \ln b \, ,\ee
where $\bar{\omega}_a$ is the pull-back to $M$ of
$-\frac{1}{2}\nb_b \nabla_a \lb^b$. Hence, using the fact that
$\fie^a$ is divergence-free on $S$, we obtain:
\be {J}_S^{\fie} = -\frac{1}{8\pi G} \oint_{S} \bar{\omega}_a
\fie^a\, d^2V \ee
Since the integrand is smooth, the future limit of $J^\fie_S$ as
we approach $S_0$ is obtained by just evaluating the right side on
$S_0$. This is precisely the angular momentum $J^\fie_\Delta$
associated with the non-expanding horizon \cite{abl2}. Thus, the
angular momentum defined on $H$ matches smoothly with that defined
on $\Delta$.

Let us consider the energy. $E^{t_o}_S$ on any cross-section $S$
of $H$ and $E^{t_0}_\Delta$ on $\Delta$ are functions of the
angular momentum and area and their form is determined by the
functional dependence on area and angular momentum of the mass
function in Kerr metrics. Since the area and angular momentum
match smoothly, we are immediately led to the conclusion that
$E^{t_o}_S$ matches smoothly with $E^{t_0}_\Delta$ as $S$
approaches $S_0$.

Finally, we will show that the vector field $t_o^a$ on $H$ matches
smoothly to a `permissible' vector field $t_0^a$ on $\Delta$.
While these fields do not have a direct physical significance as
far as final results are concerned, since energy $E^{t_o}_S$ on
$H$ is associated with the vector field $t_o^a$ and
$E^{t_0}_\Delta$ is associated with $t^a_0$, for conceptual
completeness, we need to verify that the two vector fields match
at $S_0$. To explore the relation between the two, the key idea is
to use the property
\be {\cal L}_{\Vb}\, \tilde\epsilon_{ab} = -\frac{1}{2}\, b^2
\Theta_{(\nb)} \tilde\epsilon_{ab} \ee
on $H$, where $\tilde\epsilon_{ab}$ is the intrinsic area 2-form
on the cross-sections $S$. Integrating this equation on any one
cross-section, using $a = 4\pi R^2$, and the fact that $dr/dR = 2R
\k_r$, we obtain:
\be dr = - \, \left[\frac{R\k_r}{4\pi R}\,\, \oint_S b^2
\Theta_{(\nb)} d^2V \right]\, dv \ee
on $H$ where $v$ is the affine parameter along $\Vb^a$ which takes
constant values on cross-sections on $H$. Since $N_r = |dr|$ and
$1/2b = |dv|$, it now follows that
\be  \frac{N_r}{b} + \frac{R\k_r}{8\pi R b^2}\,\, \oint_S b^2
\Theta_{(\nb)} d^2V \, = 0\, . \ee
Now, the second term admits a smooth limit to $S_0$, whence the
limit of $N_r/b$ is also well-defined. (Furthermore, since $b$ is
positive and $\Theta_{(\nb)}$ negative on $H$, it follows that the
limit is nowhere zero.) This in turn implies that $t_o^a = N_r
\ell^a -\Omega_o \fie^a$ also admits a well-defined limit to
$S_0$.

Our final task is to show that this vector field admits a smooth
extension to a `permissible' vector field $t^a_0 = {\lb_0}^a -
\Omega_0 \fie^a$ on $\Delta$, where ${\lb_0}^a$ has Kerr surface
gravity $\kappa_o$ and $\Omega_0$ is the Kerr angular velocity.
Since $N_r \ell^a$ has a smooth, nowhere vanishing limit, one can
always use the rescaling freedom in (\ref{rescaling1}) and
(\ref{rescaling2}) to choose the desired $\lb_0^a$. The matching
of $\Omega_0$ is guaranteed simply by setting $\Omega_0 =
\Omega_o$. Thus, there is a smooth vector field on $M$ which is a
`permissible' evolution field on $\Delta$ and agrees with the
`canonical' $t^a_0$ on $H$. Furthermore, this construction
provides us with a `canonical' weakly isolated horizon structure
$[\lb_0]$ on $\Delta$.

Thus, the results for transition to equilibrium at a finite time
can be summarized as follows. Assuming that $M = H\cup \Delta$ is
$C^{k+1}$, and the rotational vector field $\fie^a$ on $M$ is
$C^k$, one finds that: i) There is a $C^k$ matching of the angular
momentum $J^\fie_S$ on $H$ with $J^\fie_\Delta$ on $\Delta$; ii)
there is a unique weakly isolated horizon structure $[\lb_0]$ on
$H$ such that the canonically chosen vector field $t_o^a$ on $H$
has a $C^k$ matching with $t_0^a$ on $\Delta$; and, iii) the
corresponding energies $E^{t_o}_S$ and $E^{t_0}_\Delta$ match in a
$C^k$ manner.

If the horizon reaches equilibrium only asymptotically, $M$ is
space-like everywhere and becomes null only asymptotically. In
this case, we can just use the structure we already have on $H$
from sections \ref{s3} to \ref{s5} to describe dynamics. The
discussion of this section implies that the asymptotic state
should be identified with the weakly isolated horizon $(\Delta,
[\ell])$ where the equivalence class $[\ell]$ is determined by
$t_o^a$ on $H$. With this identification, the asymptotic limit is
reached smoothly.

We conclude with two remarks.
\begin{description}

\item[i.] \emph{surface gravity:} Following Booth and Fairhurst
\cite{bf1}, one can define surface gravity of $\Vb$ on $H$
\be \kappa_{\Vb} := - \frac{1}{2}\,\nb_b \Vb^a \nabla_a \Vb^b \,
\ee
which matches smoothly with the surface gravity $\kappa_o$ on
$\Delta$. Furthermore, one can use it to restrict the freedom in
the choice of $\Vb^a$ considerably by requiring that the 2-surface
average of $\kappa_{\Vb}$ be the effective surface gravity
$\bar\kappa_r = \kappa_o$ introduced in section \ref{s5.1}:
\be \label{average} \oint_S \kappa_V d^2V =  \kappa_o(R, J(R))
\, a_S
\ee
where $a_S$ is the area of the cross-section $S$. Then the only
remaining freedom is that of a constant:
\be \Vb^a \longrightarrow (1 + c \exp\, -\frac{1}{2}\int\kappa_o
dv \,)\, \Vb^a \ee
where $v$ is the affine parameter along $\Vb$ and $c$ a constant.
In the case when equilibrium is reached only asymptotically, all
these vector fields tend to the same null vector.

\item[ii.] \emph{Slowly evolving horizons:} Heuristically, one
expects that near the transition surface $S_0$, $H$ would become
weakly dynamical and can be regarded as a perturbed, non-expanding
horizon. However, strictly, weakly dynamical horizons can and
should be defined in their own right because, as Appendix \ref{a1}
shows, a dynamical horizon can have strong time dependence
arbitrarily close to the transition surface. A notion of `slowly
evolving horizons' has already been introduced in \cite{bf1} and
our introduction of the vector field $\Vb^a$ was motivated by that
analysis. However, in the calculation of energy, \cite{bf1} uses
only a \emph{space-like} vector field analogous to $\Vb^a$, in
place of our null vector $N_r\ell^a$. Hence that analysis is based
\emph{only} on the momentum constraint (\ref{momconstr}); the
Hamiltonian constraint (\ref{hamconstr}) plays no role. This is
probably because $\Vb^a = \lb^a - b^2 \nb^a$ and, in the leading
order approximation studied in \cite{bf1}, the $\nb^a$ term can be
neglected. However, the detailed relation between our discussion
of passage to equilibrium and that discussion of slowly evolving
horizons is yet to be understood.

\end{description}

\section{Discussion}
\label{s7}

Let us begin with a brief summary. A dynamical horizon is a
space-like 3-manifold, foliated by 2-dimensional closed,
marginally trapped surfaces $S$ (called cross-sections) on which
the expansion of the inward null normal is negative. While the
definition is so simple and conditions in it appear to be quite
weak, dynamical horizons turned out to have remarkable properties.
Specifically, we were able to: i) propose a definition of the flux
of gravitational energy falling across a portion $\Delta H$ of $H$
bounded by two cross-sections  and show that it is local,
manifestly positive and gauge invariant (i.e. does not depend on
any structure that is not intrinsic to the problem); ii) provide a
detailed area balance law relating the change in the area of $H$
to the flux of energy across it; iii) show that the cross-sections
$S$ of $H$ have the topology $S^2$ if the cosmological constant
$\Lambda$ is positive and of $S^2$ or $T^2$ if it is zero. The
$T^2$ case is degenerate in the sense that matter as well as
gravitational energy fluxes vanish, the intrinsic metric on each
cross section is flat, the shear of the (expansion-free) null
normal $\ell^a$ vanishes and the derivatives of the expansion
along both null normals vanish; iv) introduce the notion of
angular momentum associated with each cross-section and of the
flux across portions $\Delta H$ of the horizon bounded by two
cross-sections; v) provide an integral generalization of the first
law of black hole mechanics to fully dynamical situations; vi) For
axi-symmetric horizons, give a prescription to find a vector field
$t_o^a$ on $H$ and introduce a notion of energy $E^{t_o}_S$ for
each cross section $S$ such that an easily interpretable balance
law holds: if the portion $\Delta H$ of the horizon is bounded by
$S_2$ and $S_1$, then  $E^{t_o}_{S_2} -E^{t_o}_{S_1} =
\mathcal{F}^{t_o}_{\Delta H}$, where $E^{t_o}_S$ and the flux
$\mathcal{F}^{t_o}_{\Delta H}$ are both local and have physically
attractive properties; and, vii) analyze in detail the passage to
equilibrium during which a dynamical horizon becomes a weakly
isolated one.

Let us highlight a few features of the framework and the results:
\begin{itemize}
\item i) Our analysis is \emph{not} motivated by nor directly
related to the issue of finding quasi-local mass in general
relativity. Our results pertain to very special 2-surfaces ---the
cross-sections of dynamical horizons--- and cannot be applied in
more general context. Nonetheless, there is a general expectation
that a dynamical black hole space-time would admit a large number
of dynamical horizons and it is somewhat surprising that they all
have such nice properties.
\item ii) While the definition of the dynamical horizons assumes that
the expansion $\Theta_{(n)}$ of the inward pointing null normal
should be negative, most of the detailed results go through also
in the case when $\Theta_{(n)} \le 0$. Under the stronger
assumption the area monotonically increases. Under the weaker
assumption we only know that it cannot decrease, but the balance
laws and the generalization of black hole mechanics still goes
through. If $\Theta_{(n)} \ge 0$, we are in the white hole
situation in which the results again apply with appropriate sign
changes.
\item iii) The preferred vector field $t_o$ has been chosen with
the physical problems of black hole formation and coalescence in
mind. In particular, the energy $E^{t_o}_S$ associated with a
cross-section $S$ is \emph{precisely} the mass of the Kerr
space-time which has the same area and angular momentum as $S$;
one regards $S$ as being `instantaneously Kerr'. The surprising
fact is that even in the fully dynamical regime, the difference
between energies associated with two cross-sections is given by a
\emph{local}, geometrically defined flux. In the non-rotating
case, $E^{t_o}$ reduces to the Hawking mass. But in the rotating
case, the Kerr mass seems to be a better measure of the physical
energy of the horizon (associated with the vector field $t_o^a$).
\item iv) What is the situation in higher dimensions? Since our
results stemmed from the constraint part of Einstein's equations
in the metric variables, the method is directly applicable also in
higher dimensions. But the form of results would be different. In
particular, since the topological restriction made a crucial use
of the Gauss-Bonnet theorem, it will not go through; since the
black hole uniqueness theorem fails, there will be many distinct
preferred vector fields $t^a_o$ and the most convenient choice
will be dictated by the isolated horizon to which the dynamical
one settles down to; and, some of the equations may now acquire
Weyl tensor terms.
\end{itemize}
Finally, these results open up new avenues for further research in
numerical, mathematical and quantum relativity. We will conclude
by pointing out some of these.
\begin{description}
\item[\emph{Numerical and mathematical relativity:}]  In a
gravitational collapse or a black hole merger, one expects the
dynamical horizon in the distant future to asymptotically approach
a weakly isolated horizon. Can one establish that this expectation
correct? If so, what can one say about the rate of approach? While
this issue can be studied analytically, numerical simulations
provide an ideal setting to analyze it because the world tube of
apparent horizons arises there naturally and provides the
dynamical horizon. There exists a simple, local characterization
of the Kerr isolated horizon \cite{lp}. Under what conditions is
one guaranteed that the asymptotic isolated horizon is the Kerr
horizon? On an isolated horizon one can define multipoles
invariantly \cite{aaentropy} and the definition can be carried
over to each cross-section of the dynamical horizon. What can one
say about the rate of change of these multipoles? For example,
from the knowledge of the horizon quadrupole and its relation to
the Kerr quadrupole, can one gain insight in to the maximum amount
of energy that can be emitted in gravitational radiation? Is the
quasi-normal ringing of the final black hole coded in the rate of
change of the horizon multipoles, as was suggested by somewhat
heuristic considerations in the early numerical simulations
\cite{earlynr} of non-rotating black holes?
\item[\emph{Geometric analysis:}] Since $H$ is space-like, one
can consider the standard initial value problem on it. Can one
characterize the solutions to the constraint equations such that
$(H, q_{ab}, K_{ab})$ is a dynamical horizon? (It is trivial to
check that the data cannot be time symmetric (i.e. $K_{ab}$ cannot
be zero on $H$) but one could consider the constant mean curvature
case.) A full characterization would provide a complete control on
the geometry of the world tube of apparent horizons that will
emerge in \emph{all possible} numerical simulations! One can
further ask: Can one isolate the freely specifiable data in a
useful way? Are these naturally related to the freely specifiable
data on weakly isolated horizons \cite{abl1}? In the spherically
symmetric case, these issues are straightforward to address and an
essentially complete solution is known. It would be very
interesting to answer these questions in the axi-symmetric case.

Another potential application is to the proof of Penrose
inequalities which say that the total (ADM) mass of space-time
must be greater than half the radius of the apparent horizon on
any Cauchy slice. In the time symmetric case (i.e., when the
extrinsic curvature on the Cauchy slice vanishes) this conjecture
was recently proved by Huisken and Ilmamen, and Bray \cite{hib}.
Our analysis provides two flows which may be potentially useful to
extend the analysis beyond the time-symmetric case. The first is
associated with the Hawking mass and was discussed in section
\ref{s3.1}: Eq. (\ref{ab3}) shows that the Hawking mass increases
monotonically along a dynamical horizon. Furthermore, one expects
that the dynamical horizon would settle down to a weakly isolated
horizon in the future. For isolated horizons which extend all the
way to $i^+$, under certain regularity conditions, the horizon
mass is the future limit of the Bondi mass \cite{abf}. Thus, using
our flow, one should be able to prove a stronger version of the
Penrose inequality where the ADM mass is replaced by the future
limit of the Bondi mass. The second flow is associated with the
Kerr mass and was discussed in section \ref{s5}. In the
non-rotating case, this is the same as the first flow. But in the
rotating case, the Kerr mass is greater than the Hawking mass
whence it would provide a further strengthening of the Penrose
inequality. Note that the Kerr mass increases monotonically only
if one begins with a cross-section $S$ of $H$ on which $2GJ <
R^2$. But if we initially violate this condition, the flow drives
the system towards satisfaction of this inequality.
\item[\emph{Quantum relativity:}] As the vast mathematical
literature on black hole mechanics shows, the infinitesimal
version (\ref{1law1}) of the first law has had a deep conceptual
influence. The finite version (\ref{integral1law}) may have
similar ramifications in non-equilibrium situations. The
Hamiltonian framework of Booth and Fairhurst \cite{bf2} could be
used as a point of departure for describing quantum black holes
beyond equilibrium situations. Can one, in particular, extend the
non-perturbative quantization of isolated horizons of
\cite{abck,aaentropy} to describe quantum, dynamical horizons? To
calculate their entropy? To naturally incorporate back reaction in
the Hawking process?
\end{description}

\section*{Acknowledgements:} We would like to thank
Ivan Booth, Stephen Fairhurst, Sean Hayward, Jerzy Lewandowski and
Tomasz Pawlowski for stimulating discussions and Lars Andersson,
Chris Beetle, Bobby Beig, Matt Choptuik, Piotr Chrusciel, Sergio
Dain, Sam Finn, Eanna Flanagan, Greg Galloway, Jim Hartle, Gary
Horowitz, Gerhard Huisken, Tom Ilmanen, Jim Isenberg, Pablo
Laguna, Luis Lehner, Bernd Schmidt, Walter Simon, Bill Unruh and
Jeff Winicour for interesting comments and questions. This work
was supported in part by the National Science Foundation grants
PHY-0090091, PHY99-07949, the National Science Foundation
Cooperative Agreement PHY-0114375, the Eberly research funds of
Penn State and the Albert Einstein Institut.

\begin{appendix}

\section{Examples: The Vaidya solutions}
\label{a1}

The Vaidya metrics provide simple, explicit examples of dynamical
horizons. Furthermore, when the flux of the null matter field
vanishes, one obtains an isolated horizon. Therefore, the metrics
also provide explicit examples of the transition from the
dynamical to isolated horizons discussed in section \ref{s6}.
In section \ref{a1.1} we describe the Schwarzschild-Vaidya dynamical
horizon and in section \ref{a1.2}, we include the cosmological
constant.

\subsection{The Schwarzschild--Vaidya metrics}
\label{a1.1}

In the ingoing Eddington-Finkelstein coordinates
$(v,r,\theta,\phi)$ the 4-metric is given by
\be g_{ab} = -\left(1-\frac{2GM(v)}{r}\right)\grad_av\grad_bv +
2\grad_{(a}v\grad_{b)}r +
r^2\left(\grad_a\theta\grad_b\theta+\sin^2\theta\grad_a\phi\grad_b\phi\right)
\ee
where $M(v)$ is any smooth non-decreasing function of $v$. This is
a solution of Einstein's equations with zero cosmological
constant, the stress-energy tensor $T_{ab}$ being given by
\be T_{ab} = \frac{\Mdot(v)}{4\pi r^2}\grad_av\grad_bv \ee
where $\Mdot=dM/dv$. Clearly, $T_{ab}$ satisfies the dominant
energy condition if $\Mdot \geq 0$ and vanishes if and only if $\Mdot =0$.

Let us focus our attention on the metric 2-spheres given by $v
={\rm const}, r= {\rm const}$. The outgoing and ingoing null
normals to these 2-spheres can be taken to be, respectively,
\be \bar\ell^a = \left(\frac{\partial}{\partial v}\right)^a +
\frac{1}{2}\left(1-\frac{2GM}{r}\right)
\left(\frac{\partial}{\partial r}\right)^a \,
\quad {\rm and} \quad \bar{n}^a = -2 \left(\frac{\partial}{\partial
r}\right)^a
\ee
(so that $\bar{\ell}^a \bar{n}_a = -2$ as in the main text). The expansion of
the outgoing null normal $\bar\ell^a$ is given by:
\be \Theta_{(\bar\ell)} = \frac{r-2GM(v)}{r^2} \, . \ee
Thus, the only spherically symmetric marginally trapped surfaces
are the 2-spheres $v ={\rm const}$ and $r = 2GM(v)$. The question
is if these surfaces are cross-sections of a dynamical horizon. On
each of these surfaces, the expansion of the ingoing normal $n^a$
is negative, $\Theta_{(\bar{n})} = -4/r$. Furthermore, at the
marginally trapped surfaces, $\Lie_{\bar{n}}\theta_{(\bar{\ell})}
= -2/r^2 < 0$. Because of spherical symmetry the shear of
$\bar{\ell}^a$ (and $\bar{n}^a$) vanishes identically. Finally,
$T_{ab}\bar{\ell}^a\bar{\ell}^b > 0$ if and only if $\Mdot > 0$.
Hence it follows from our general discussion in section \ref{s2.1}
(see equation (\ref{spacelike})) that the surface $H$ given by
\be r=2GM(v)\quad \textrm{with} \quad \Mdot > 0 \ee
is a dynamical horizon. When $\Mdot$ vanishes, the surface $r =
2GM(v)$ becomes null and a non-expanding horizon (which is, in
fact, an isolated horizon). The full surface $r =2GM(v)$ is a
future outer trapping horizon (FOTH) of Hayward's \cite{sh}.

The null normals $\bar{\ell}^a$ and $\bar{n}^a$ are well suited
for studying the approach to equilibrium, i.e., the transition
from the dynamical to the isolated horizon discussed in section
\ref{s6}. The interpolating vector field $\Vb^a$ is now given by:
\be \Vb^a = \left(\frac{\partial}{\partial v}\right)^a + 2 G\Mdot\,
\left(\frac{\partial}{\partial r}\right)^a \equiv \lb^a - G\Mdot\, \nb^a \ee
so that $b^2 = G\Mdot$. The degree of smoothness of the matching
between the isolated and the dynamical horizons is dictated by
differentiability of $\Mdot$ at $S_0$. Thus, if $\Mdot$ is $C^k$
on $S_0$, physical fields will match in a $C^k$ fashion. Finally,
because of spherical symmetry the surface gravity $\kappa_{\Vb}$
of $\Vb^a$ is constant on each cross-section and is given by
$1/2R(v)$; the canonical vector field $t_0^a =
(\partial/\partial\, v)^a$ on $\Delta$ matches smoothly with the
canonical vector field $t_o^a = (\partial/\partial\, v)^a$ on $H$;
the angular momentum vanishes; and, the horizon mass is given by
$M(v)$.

To study the structure of $H$ by itself, as in the main body of
the paper, a different normalization of null vectors is more
convenient. For completeness, we list all the relevant vector
fields:
\ba \hat{\tau}_a =
\frac{1}{2\sqrt{G\Mdot}}\grad_ar-\sqrt{G\Mdot}\grad_av &\qquad&
\hat{r}_a = \frac{1}{2\sqrt{G\Mdot}}\grad_ar+\sqrt{G\Mdot}\grad_av
\nonumber\\
n_a = -2\sqrt{G\Mdot}\grad_a v &\qquad&
\ell_a = \frac{1}{\sqrt{G\Mdot}}\grad_a r \nonumber\\
\hat{r}^a = \frac{1}{2\sqrt{G\Mdot}} \left(\frac{\partial}{\partial
v}\right)^a + \sqrt{G\Mdot}\left(\frac{\partial}{\partial
r}\right)^a &\qquad& \hat{\tau}^a =
\frac{1}{2\sqrt{G\Mdot}}\left(\frac{\partial}{\partial v}\right)^a
- \sqrt{G\Mdot}\left(\frac{\partial}{\partial r}\right)^a
\nonumber\\
n^a = -2\sqrt{G\Mdot}\left(\frac{\partial}{\partial r}\right)^a
&\qquad& \ell^a = \frac{1}{\sqrt{G\Mdot}}
\left(\frac{\partial}{\partial v}\right)^a\, . \ea
Let us take $(r,\theta,\phi)$ as coordinates on the dynamical
horizon. The radial coordinate $r$ is also the area coordinate $R$
in this case, whence $N_R$ and $\k_R$ of the main text will be
denoted just by  $N_r$ and $\k_r$ respectively. $N_r$ is given
simply by
\be N_r = \sqrt{G\Mdot}\, . \ee
Therefore, the matter flux is
\be \mathcal{F}^{(r)}_\m = \int_{\Delta H}
N_rT_{ab}\hat{\tau}^a\ell^b\,d^3V = \frac{1}{8\pi G}\int_{\Delta
H} \frac{1}{r^2}\,dr\,d^2V = \frac{r_2-r_1}{2G} \, . \ee
The gravitational flux, of course, vanishes because of spherical
symmetry.

\emph{Remark:} In the above discussion, we restricted ourselves to
dynamical horizons whose cross-sections are spherically symmetric.
It is natural to ask if the space-time admits other, non-spherical
dynamical horizons. Surprisingly, this question is not easy to
analyze because very little is known about non-spherical
marginally trapped surfaces even in the Schwarzschild space-time.
However, we will show that, within the $v= {\rm const}$ surfaces,
there is no marginally trapped 2-surface which lies entirely
outside the surface $r = 2M(v)$ considered here. It would be
interesting to know if the space-time admits other dynamical
horizons and, if so, whether the one discussed here is Hayward's
trapping boundary \cite{sh}, discussed in section \ref{s2.2}.

Let us then look for a closed 2-surface $S^\prime$ given by $r =
2GM(v)-h(\theta,\phi)$ which is marginally trapped. By
construction, it lies on the constant $v$ slices. Let
$\tilde{\ell}^a$ and $\tilde{n}^a$ be null normals to this
2-surface. One can show that $\tilde{n}^a = \bar{n}^a$ but
$\tilde{\ell}^a \neq \bar{\ell}^a$.  It can also be shown that the
ingoing null expansion to is still $\Theta_{(\tilde{n})} = -4/r$
while the outgoing expansion becomes
\be
\Theta_{\tilde{(\ell)}} = -\frac{h}{r^2} + \frac{\Delta_0 h}{r^2} +
\frac{|D_0 h|^2}{r^3}
\ee
where $\Delta_0$ and $D_0$ are respectively the standard Laplacian
and derivative operator on the unit 2-sphere in $(\theta,\phi)$
coordinates.   By setting $\Theta_{(\tilde{\ell})}=0$ we obtain
the following partial differential equation for $h(\theta,\phi)$:
\be \label{vaidyamts}\Delta_0 h - h = -\frac{|D_0 h|^2}{r} \, .\ee
As expected, $h=0$ is clearly a solution; but is it the unique
solution? Integrate both sides of eq. (\ref{vaidyamts}) using the
standard unit 2-sphere volume element and obtain the inequality
$\oint_{S^\prime} h > 0$.  This tells us that we cannot have
solutions to eq. (\ref{vaidyamts}) with $h$ everywhere negative.
In other words, we cannot have marginally trapped surfaces which
lie completely outside $r=2GM(v)$.  Of course, the analysis is
incomplete because it does not preclude surfaces which lie only
partially outside $r=2GM$ nor surfaces which do not lie on the
$v=\textrm{const}$ slices; these issues are currently under
investigation.

\subsection{Inclusion of the cosmological constant}
\label{a1.2}

The example presented in the last sub-section can be generalized
to include a cosmological constant $\Lambda$.  For definiteness, we
restrict ourselves the $\Lambda>0$ case. The Vaidya metric in the
presence of a cosmological constant is
\be g_{ab} = -\left(1-\frac{2GM(v)}{r}-\frac{\Lambda
r^2}{3}\right)\grad_av\grad_bv + 2\grad_{(a}v\grad_{b)}r +
r^2\left(\grad_a\theta\grad_b\theta+
\sin^2\theta\grad_a\phi\grad_b\phi\right) \, . \ee
As before, $M(v)$ is a non-decreasing function of $v$.  When $M$
is a constant, this is just the usual Schwarzschild-de Sitter
solution. As we shall see below, this solution admits a black hole
horizon only if the inequality $9\Lambda (GM)^2 \leq 1$ is
satisfied.  In the remainder of this section we shall always
assume that $GM$ never exceeds the value $(9\Lambda)^{-1/2}$. The
Einstein tensor for the metric given above is
\be G_{ab} = -\Lambda g_{ab} + \frac{2G\Mdot}{r^2} \grad_av\grad_bv
\, . \ee
As before, the stress energy tensor $T_{ab}$ is
\be T_{ab} = \frac{\Mdot(v)}{4\pi r^2}\grad_av\grad_bv \ee
and $\Mdot \geq 0$ is required in order to satisfy the null
energy condition.
In this case, there are two horizons: the usual black hole horizon and
also a cosmological horizon which are given by the solutions of the equation
\be \label{eq:cubic} f(v,r) := 1-\frac{2GM(v)}{r}-\frac{\Lambda r^2}{3} =
0 \, . \ee
This is a cubic equation in $r$ and when $0 < 9\Lambda (GM)^2 < 1$,
it admits precisely two real and positive solutions given by
\be r_c = \frac{2}{\sqrt{\Lambda}} \cos\left(\frac{\pi
-\alpha}{3}\right) \qquad \textrm{and} \qquad r_b =
\frac{2}{\sqrt{\Lambda}} \cos\left(\frac{\pi + \alpha}{3}\right)
\ee
where $\alpha = \cos^{-1}(\sqrt{9\Lambda (GM)^2})$.  The black
hole horizon is located at $r_b$ and the cosmological horizon at
$r_c$. In general, $r_b \leq r_c$ and when $9\Lambda (GM)^2 = 1$,
the two horizons coincide: $r_b = r_c = \sqrt{3/\Lambda}$.  When
$9\Lambda (GM)^2 >1$, then equation (\ref{eq:cubic}) does not
admit any real positive solutions. Assuming that $M$ is an
increasing function of $v$, it is easy to see that $r_b$ increases
with time and $r_c$ decreases with time and both horizons merge in
the limit $GM^2 \rightarrow 1/9\Lambda$.

The derivatives of $f(v,r)$ are
\be f^\prime = \frac{\partial f}{\partial r} = \frac{2GM}{r^2} -
\frac{2\Lambda r}{3} \qquad \textrm{and} \qquad \dot{f} =
\frac{\partial f}{\partial v} = -\frac{2G\Mdot}{r} < 0 \ee
At the horizons, when $f=0$, the expression for $f^\prime$ simplifies
to
\be f^\prime|_{f=0} = \frac{1-\Lambda r^2}{r} \,.\ee
The derivative $f^\prime$ is positive at the black hole
horizon and negative at the cosmological horizon.

As before, let us look for all possible spherically symmetric
marginally trapped surfaces. The null normals to the
$r=\rm{constant}$, $v=\rm{constant}$ surfaces can be taken to be
\be
\bar{\ell}^a = \left(\frac{\partial}{\partial v}\right)^a +
\frac{f}{2}\left(\frac{\partial}{\partial r}\right)^a \qquad
\textrm{and} \qquad \bar{n}^a = -2\left(\frac{\partial}{\partial
  r}\right)^a\, .
\ee
The expansions of these null normals are
\be \Theta_{(\bar{\ell})} = \frac{f}{r} \qquad \textrm{and} \qquad
\Theta_{(\bar{n})} = -\frac{4}{r} \, .\ee
Thus we see that $\Theta_{(\bar{n})}$ is always negative and
$\Theta_{(\bar{\ell})}$ vanishes precisely at the two horizons.
Furthermore,
\be \Lie_{\bar{n}}\Theta_{(\bar{\ell})}|_{f=0} = -\frac{2f^\prime}{r}
\ee
which tells us that $\Lie_{\bar{n}}\Theta_{(\bar{\ell})} < 0$ at
$r=r_b$ and $ > 0$ at $r=r_c$.  Thus, if $\Mdot > 0$,  the surface
$r=r_b$ is space-like and is a dynamical horizon while $r=r_c$ is
time-like and is, in fact, a \emph{time-like dynamical horizon} as
discussed in appendix \ref{a3}.

In the remainder of this section, we focus only on the black hole
horizon. All the remaining equations in this sub-section are valid
only at $r=r_b$.  The unit normal to the horizon is
\be \hat{\tau}_a =
\frac{1}{\sqrt{|2\dot{f}f^\prime|}}\left[\dot{f}\grad_av +
f^\prime\grad_ar\right] \qquad \textrm{and} \qquad \hat{\tau}^a =
\frac{1}{\sqrt{|2\dot{f}f^\prime|}}\left[f^\prime\left(\frac{\partial}{\partial
v}\right)^a + \dot{f}\left(\frac{\partial}{\partial
r}\right)^a\right] \, . \ee
The constant $r$ surfaces are the preferred cross sections of the
horizon and the unit space-like normal $\hat{r}^a$ to these cross
sections is
\be \hat{r}_a =
\frac{1}{\sqrt{|2\dot{f}f^\prime|}}\left[-\dot{f}\grad_av +
f^\prime\grad_ar\right] \qquad \textrm{and} \qquad \hat{r}^a =
\frac{1}{\sqrt{|2\dot{f}f^\prime|}} \left[f^\prime\left(
\frac{\partial}{\partial v}\right)^a
 - \dot{f}\left( \frac{\partial}{\partial r}\right)^a \right]
\ee
The properly rescaled null normals are
\be \ell^a = \frac{2|f^\prime|}{\sqrt{|2\dot{f}f^\prime|}} \left(
\frac{\partial}{\partial v}\right) ^a \qquad \textrm{and} \qquad
n^a = \frac{2\dot{f}}{\sqrt{|2\dot{f}f^\prime|}} \left(
\frac{\partial}{\partial r}\right) ^a  \, . \ee
The lapse function corresponding to the radial coordinate $r$, which
in this case is also the area radius, is given by
\be N_r = \left|\frac{\dot{f}}{2f^\prime}\right|^{1/2} =
\sqrt{\frac{G\dot{M}}{|1-\Lambda r^2|}} \ee
and thus the properly rescaled vector field corresponding to the
radial coordinate $r$ is $t_o^a = N_r\ell^a = (\partial /\partial
v)^a$.

To calculate the flux law, let us first compute
$T_{ab}\hat{\tau}^a\ell^b$:
\be T_{ab}\hat{\tau}^a\ell^b = \left(\frac{2\dot{M}}{r^2}\right)
\frac{|f^\prime|}{\sqrt{|2\dot{f}f^\prime|}} \cdot
\frac{2|f^\prime|}{\sqrt{|2\dot{f}f^\prime|}} = \frac{f^\prime}{r}
= \frac{1-\Lambda r^2}{r^2}\, . \ee
Therefore, the matter flux across the dynamical horizon is:
\be \mathcal{F}^{(r)}_{\m} = \int_{\Delta H} N_r
T_{ab}\hat{\tau}^a\ell^b\,d^3V = \frac{1}{8\pi G}\int_{\Delta H}
\left(\frac{1}{r^2}-\Lambda\right) \, dr\,d^2V \ee
and the mass function on the horizon is
\be E^{t_0}(r) = \frac{r}{2G} - \frac{\Lambda r^3}{6G} = M(v)\, ,
\ee
whence, as expected, the infinitesimal form of the first law takes
the form (\ref{1law5}).

\section{Time-like analogs of dynamical horizons}
\label{a3}

In the analysis presented in this paper, the space-like character
of the dynamical horizon played a crucial role.  However, as we
saw in appendix \ref{a1.2}, the time-like case can occur in
cosmological contexts. We do not expect the matter or the
gravitational fluxes to be generally positive definite in this
case whence, in particular, the topology of cross-sections need
not be restricted. However, for simplicity of presentation, we
shall consider only the case of spherical topology; the
generalization to higher genus cross-sections is obvious.

\noindent \textbf{Definition:} A smooth, three-dimensional,
time-like sub-manifold $H$ in a space-time is said to be a
\emph{time-like dynamical horizon} if it is foliated by a family
of space-like 2-spheres such that on each leaf, the expansion
$\theta_\ls$ of a null normal $\l^a$ vanishes while the expansion
$\theta_\ns$ of the other null normal $n^a$ is strictly negative.

The notation will follow the space-like case as much as possible.
The main difference is that $\rh_a$ and $\th_a$ now have different
meanings. $\rh_a$ is no longer tangential to $H$, it is instead
the unit space-like vector \emph{normal} to $H$.  Similarly,
$\th_a$ is the unit time-like vector \emph{tangential} to $H$ and
orthogonal to the cross-sections of $H$.  As before, the null
normals are
\be \l^a = \th^a + \rh^a \qquad \textrm{and} \qquad n^a = \th^a -
\rh^a \, . \ee
As one would expect, the time-like case involves many quantities
which are analogues of their space-like counterparts, usually with
$\rh_a$ and $\th_a$ interchanged.  These quantities will be
denoted with primes.

What happens to the area increase law now?  Looking at the
expressions for the expansions of the null normals, one can easily
check that
\be D_a\th^a = \theta_\ls + \theta_\ns < 0 \,\, . \ee
This clearly shows that the area of the cross-sections
\emph{decreases} along $\th^a$.

As in the space-like case, the analysis of the flux law will be
based on the constraint equations on $H$.  In the time-like case,
the only difference in the constraint equations is a sign change
in the scalar constraint (compare with eqs. (\ref{hamconstr}) and
(\ref{momconstr})):
\ba {H}^\prime_S &:=& -\R + K^2 - K^{ab}K_{ab}= 16\pi G
\T_{ab}\rh^{\,a}\rh^{\,b}
\label{eq:timehamconstr}\\
{{H}^\prime_V}^a &:=& D_b\left(K^{ab} - Kq^{ab}\right) = 8\pi G
\T^{bc}\rh_{\, c}{q^a}_b \label{eq:timemomconstr} \, , \ea
where, as in the main text, $\T_{ab}$ is related to the matter
stress-energy $T_{ab}$ via $\T_{ab} = T_{ab} - (1/8\pi G) \Lambda
g_{ab}$. Once again, we focus our attention on the energy flux
along the vector $\xi^a_{(t)} = N_t\l^a$ where the lapse function
$N_t$ is now tied to the choice of a time coordinate $t$ on $H$ by
the equation
\be D_at = N_t\th_a \ee
where level surfaces of $t$ are the cross sections of $H$ and we
have the same rescaling freedom in the lapse as before. The
expression for the matter energy flux along $\xi_{(t)}^a$ is now
given by
\be \mathcal{F}^{(t)}_\m:= \int_{\Delta H}
\T_{ab}\rh^{\,a}\xi_{(r)}^b d^3V =\frac{1}{16\pi G} \int_{\Delta
H}\, N_t\left(-\R + K^2 - K^{ab}K_{ab} + 2\th_aD_bP^{ab}\right) \,
d^3V \, , \ee
the only difference from the space-like being the different sign
of the scalar curvature term.

Using the Gauss-Codacci equation relating the curvatures of $H$
and $S\subset H$ leads to
\be -\R = 2(\R_{ab} - \G_{ab})\th^a\th^b = -\twoR
+\K^2-\K_{ab}\K^{ab}+ 2D_a{\alpha^\prime}^a \ee
where
\be {\alpha^\prime}^a = \th^bD_b\th^a - \th^aD_b\th^b\, . \ee
The momentum constraint is unchanged and so the flux becomes
\be \mathcal{F}^{(t)}_\m := \frac{1}{16\pi G} \int_{\Delta H}\,
N_t\left(-\twoR + \K^2 - \K^{ab}\K_{ab} + K^2 - K_{ab}K^{ab}
-2P^{ab}D_a\th_b + 2D_a{\gamma^\prime}^a\right) \, d^3V \, , \ee
where ${\gamma^\prime}^a= {\alpha^\prime}^a+{\beta^\prime}^a$ and
${\beta^\prime}^a = K^{ab}\th_b-K\th^a$.

The decomposition of the extrinsic curvatures $K_{ab}$ and
$\K_{ab}$ proceeds exactly as before. However, the analogues of
eqs. (\ref{eq:dnr}) and (\ref{eq:projgamma}) now have some
negative signs:
\be \th^aD_a\th_b = -\frac{\twoD_b N_t}{N_t} \qquad \textrm{and}
\qquad \gamma^a = \q_b^{\,\,a}\gamma^b = \th^bD_b\th^a - \W^a\, .
\ee
The simplified flux equation then becomes
\be \mathcal{F}^{(t)}_\m = \frac{1}{16\pi G}\int_{\Delta H}
N_t\left(-\twoR - |\sigma|^2 + 2|\zeta^\prime|^2 \right)\,d^3V \ee
where ${\zeta^\prime}^a$ is the analog of $\zeta^a$:
\be {\zeta^\prime}^a = \q^{ab}\th^c\nabla_c\l_b \, . \ee
The area balance law now reads
\be \label{eq:tlbalancel} \left(\frac{R_2}{2G}-
\frac{R_1}{2G}\right) = -\int_{\Delta H}
\T_{ab}\rh^{\,a}\xi_{(t)}^b\,d^3V - \frac{1}{16\pi G} \int_{\Delta
H} N_t\left\{ |\sigma|^2 - 2|\zeta^\prime|^2\right\} \,d^3V \,
,\ee
where $\rh^a$ and $\zeta^a$ are space-like. Therefore, even though
the area decreases monotonically, neither the matter term nor the
geometrical terms on the right side have definite signs. This is
yet another illustration of the fact that the form of the area
balance law (\ref{ab1}) for dynamical horizons is very special.

\end{appendix}

\end{document}